\documentclass[11pt]{article}
\usepackage{times}
\usepackage{geometry}
\usepackage[round]{natbib}
\usepackage[utf8]{inputenc}
\usepackage[T1]{fontenc}
\usepackage{hyperref}
\usepackage{url}
\usepackage{booktabs}
\usepackage{amsfonts}
\usepackage{amsmath}
\usepackage{nicefrac}
\usepackage{microtype}
\usepackage{xcolor}
\usepackage{graphicx}
\usepackage{subcaption}
\usepackage{placeins}
\usepackage{float}
\usepackage{authblk}

\hypersetup{
  colorlinks=true,
  linkcolor=blue,
  citecolor=blue,
  urlcolor=blue
}

\begin{document}

% --- Title ---
\title{Retrieval and competition: how a protein foundation model starts a protein}

\author[1]{Piotr Jedryszek}
\author[2, 3]{Oliver M. Crook  \thanks{\url{oliver.crook@chem.ox.ac.uk}~}}

\affil[1]{Department of Biology, University of Oxford, Oxford, UK}
\affil[2]{Kavli Institute for Nanoscience Discovery,
	University of Oxford, Oxford, UK}
\affil[3]{Department of Chemistry, University of Oxford, Oxford, UK}

\date{}
\maketitle

% =============================================================================
% ABSTRACT
% =============================================================================
\begin{abstract}

Protein language models are increasingly used to guide experimental and clinical decisions, yet it is often unclear whether a confident prediction reflects recognition of biological evidence or retrieval of a statistical default. We examine this distinction for a simple near-universal biological rule, that proteins begin with methionine, by tracing the full computational pathway through which ESM2-8M, an 8 million parameter protein language model, produces this prediction. The model does not detect methionine at the masked position. Instead, it retrieves a methionine-favouring signal from a reference representation at the beginning-of-sequence token, using a position-specific query assembled across multiple layers. The final prediction emerges through competition with context-dependent circuits. To understand how positional information reaches the readout, we introduce a norm–direction decomposition of attention scores within rotary frequency bands. This reveals that positional encoding operates through coupled changes in query-norm and angular alignment, distributed across rotary frequency bands: the dominant short-period band engages both factors, while longer-period bands contribute primarily through angular alignment. On sequences whose true N-terminus is not methionine, precisely the cases where the biological question matters, the model predicts methionine anyway. The output is not a correct prediction produced by an unexpected mechanism; it is the output of a positional-prior retrieval circuit that happens to match the statistical average and fails exactly where biology diverges from that average. Distinguishing the two requires resolution at the level of individual circuits, frequency bands, and query composition, suggesting that mechanistic verification will be necessary, and challenging, for predictions where the biological stakes are higher. Even for the simplest biological rule, the model’s prediction is mediated by a distributed computational circuit rather than direct recognition, suggesting that increasing task complexity will further obscure the relationship between model confidence and underlying biological evidence.

\end{abstract}

% =============================================================================
% INTRODUCTION
% =============================================================================
\section{Introduction}

AI models trained on millions of protein sequences are now routinely used for prediction, interpretation, and design in protein science~\citep{Lin2023Evolutionary-scaleModel, Rives2021BiologicalSequences}. From predicting mutational effects~\citep{Meier2021LanguageFunction} to annotating protein function~\citep{Chen2025EvaluatingReview} and guiding the design of novel proteins~\citep{Madani2023LargeFamilies}, protein language models (PLMs) achieve strong performance despite being trained only on sequence data, suggesting that they internalise biologically meaningful structure from sequence statistics alone. Yet their predictions remain difficult to interpret mechanistically. A confident output tells us the model is certain, but not whether that certainty reflects recognition of biologically meaningful features in the input or retrieval of a stored statistical default applied independently of input-specific evidence. Both produce the same prediction; only the first provides grounds for trusting the prediction to generalise. When such predictions guide experiments, this distinction matters: confidence alone does not guarantee that the model has used the right biological evidence.

Methods that open this black box have begun to address the problem. In natural-language models, researchers have identified the internal components responsible for in-context generalisation~\citep{Olsson2022In-contextHeads}, traced the causal pathways that connect them~\citep{Marks2025SparseModels}, and isolated features that activate for specific concepts such as cities, emotional tone, or syntactic categories~\citep{TrentonBricken2023TowardsLearning}. Similar approaches are now being applied to PLMs. Probing studies show that internal representations encode secondary structure and residue contacts~\citep{Vig2020BERTologyModels}. Sparse autoencoders decompose representations into biologically interpretable features~\citep{Simon2024InterPLM:Autoencoders, Garcia2025InterpretingAutoencoders, Parsan2025TowardsAutoencoders, Maiwald2025Decode-gLM:Models}. More recent work has begun to identify the computational pathways behind specific predictions~\citep{Nainani2025MechanisticModels, Tsui2026ProteinTranscoders, Lu2026MechanismsESMFold}. Yet attention itself — the mechanism that moves information between positions inside the model — is largely treated as a black box, and the role of positional encoding in routing that information has received little scrutiny. We know which components matter, but rarely how they compute. This gap is consequential, because protein sequences are not intrinsically interpretable to humans, and PLMs now outperform human experts on several tasks, raising the possibility that they have learned biological patterns we do not yet know. Distinguishing known biology, unknown biology, and statistical shortcuts requires understanding the computation itself; and specifically, it requires resolution at the level of individual circuits: what the model attends to, what it writes, and where the information comes from.

We address this gap by focusing on one of biology's simplest and most universal rules: proteins typically begin with methionine, reflecting translation initiation at the AUG start codon (though the initial methionine is often removed post-translationally~\citep{Frottin2006TheCleavage, Giglione2004ProteinExcision}). When the first residue is masked, PLMs~\citep{Lin2023Evolutionary-scaleModel, Meier2021LanguageFunction} overwhelmingly predict methionine, even for sequences whose mature N-terminus is something else. The models have evidently learned this rule from sequence statistics alone. Of all PLM predictions, this is the one a reader would most confidently expect the model to implement through recognition: detect the position-0 identity, write methionine because it belongs there. The rule is trivial; one might expect an equally simple implementation.

\begin{figure}[htbp!]
    \centering
    \includegraphics[width=\textwidth]{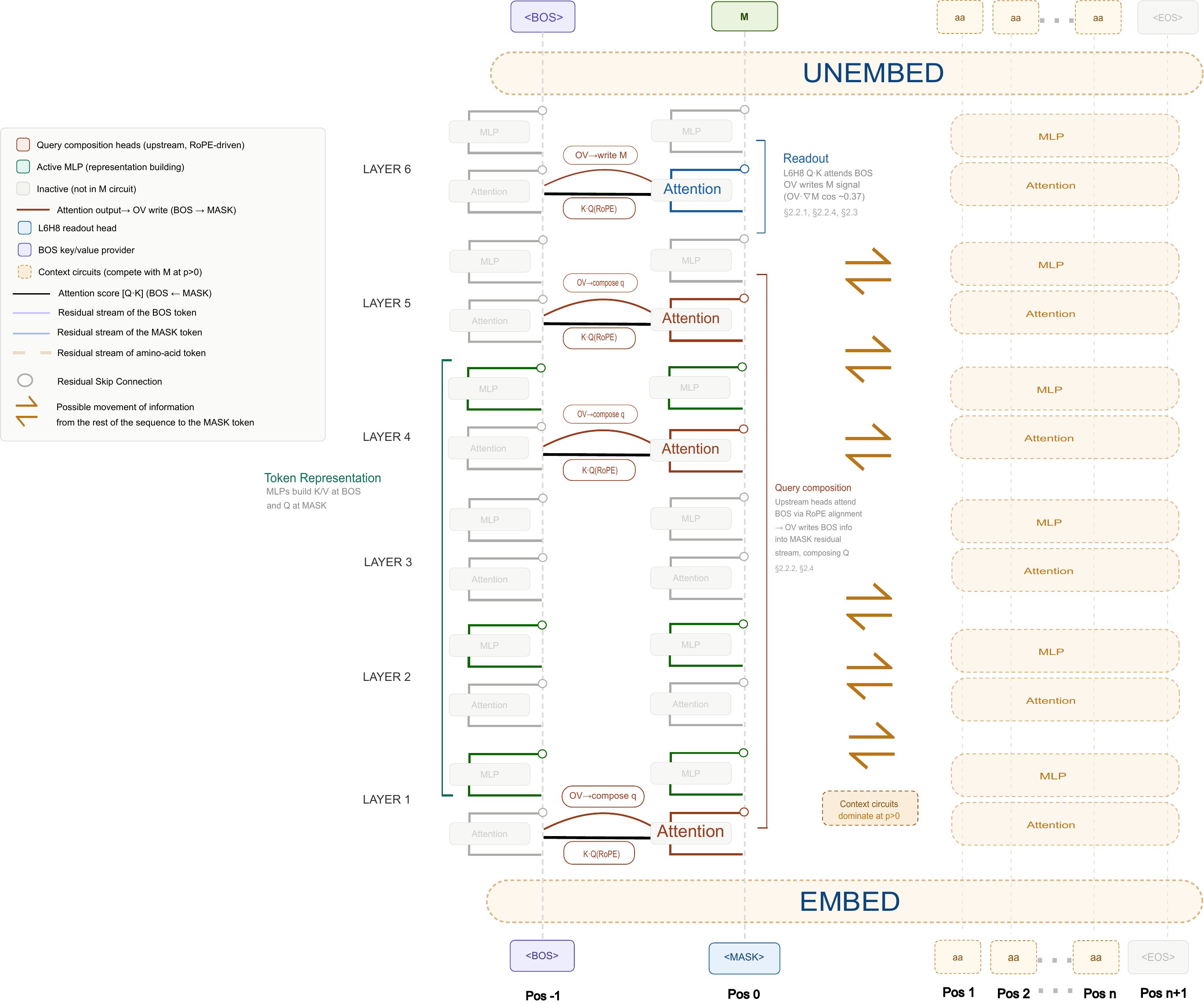}
    \caption{\textbf{Circuit diagram for positional methionine prediction in
    ESM2-8M.}
    Schematic of information flow through the methionine circuit, shown for the
    \texttt{<BOS>} token (left, purple stream) and \texttt{<MASK>} token at
    position~0 (center, blue stream). The circuit comprises three functional
    tiers. \emph{Token representation} (Layers~1--4): early-layer MLPs (green)
    progressively build token representations at BOS, which is key for correct key and value formation, and build the MASK token representation, which is crucial for correct query formation in layer 6. 
    ( \nameref{sec:circuit}).
    \emph{Query composition} (Layers~1, 4, 5): upstream attention heads
    (orange) use
    RoPE-mediated positional signals to attend to BOS and write BOS-derived
    information into the masked token's residual stream, composing the right query
    that Layer~6 will consume ( \nameref{sec:circuit},  \nameref{sec:causality}). When the RoPE is not aligned with position 0, the heads move other information into the masked token, preventing the correct query formation and attention to BOS, which underlies the Methionine circuit. 
    L4H3 is marked as context-dependent, reflecting its partial reliance on
    residual-stream content beyond RoPE (Figure~\ref{fig:patching}f). \emph{Readout} (Layer~6): L6H8 (blue) uses
    the composed query, amplified by its own RoPE rotation, to attend
    strongly to BOS (black line, Q$\cdot$K attention score) and writes a
    methionine-specific signal into the output (red arrow, OV write;
    OV$\cdot\nabla$M cosine $\sim$0.37;  \nameref{sec:circuit},
     \nameref{sec:rope}). Grey components are inactive within the minimal
    circuit. At non-initial positions ($p > 0$), context-dependent circuits
    (right, dashed tan) that process sequence tokens (``aa'') produce
    competing signals that can outweigh the positional methionine circuit,
    explaining the model's specificity without requiring explicit inhibition
    ( \nameref{sec:behaviour}).}
    \label{fig:circuit_diagram}
\end{figure}

We find the opposite. In ESM2-8M, a six-layer transformer with 8 million parameters, the rule is implemented by a distributed, multi-layer circuit — a set of interacting components that together compute the prediction — far more elaborate than the rule it encodes (Figure~\ref{fig:circuit_diagram}). The model does not store a methionine signal at the masked position itself. Instead, it maintains a stable reference representation at the beginning-of-sequence (BOS) token, a token typically discarded as a passive delimiter in encoder PLMs, which is read by a dedicated attention head in the final layer. That head only retrieves the reference when a position-specific query has been assembled upstream, across several layers, by attention heads that read out first-position identity from rotary positional embeddings (the mechanism by which transformers encode where each residue sits in the sequence). The prediction is therefore a retrieval from a shared reference, not a local rule. And it is not deterministic: the methionine circuit contributes a roughly constant signal at the first position, while parallel context-dependent circuits contribute competing predictions. The final output reflects which signal dominates, and specificity at the first position, the model's correct restraint from predicting methionine elsewhere, arises not from inhibition within the methionine circuit but from this competition between circuits.

None of these components: the active role of BOS, the distributed construction of the positional query, or the competitive readout, are visible at the level of latent features or block-level representations. They emerge only once the analysis operates at the resolution of individual attention heads, rotary frequency bands, and query composition, and we argue that this resolution will be necessary to understand how PLMs represent richer biological signals, including the ones we do not yet recognise as biology.

% --- FIGURE 1: Circuit Diagram ---
\clearpage
% =============================================================================
% RESULTS
% =============================================================================
\section{Results}

% -----------------------------------------------------------------------------
\subsection{A robust positional prior shaped by circuit competition}
\label{sec:behaviour}

Protein language models strongly prefer methionine at the first position. When the initial residue of a sequence is masked, ESM2-8M predicts methionine across a wide range of inputs, including sequences whose true N-terminal residue is not methionine. This behaviour is consistent with the model internalising a simple positional rule: whatever appears first is likely to be methionine.

\begin{figure}[htbp!]
    \centering
    \includegraphics[width=0.9\textwidth]{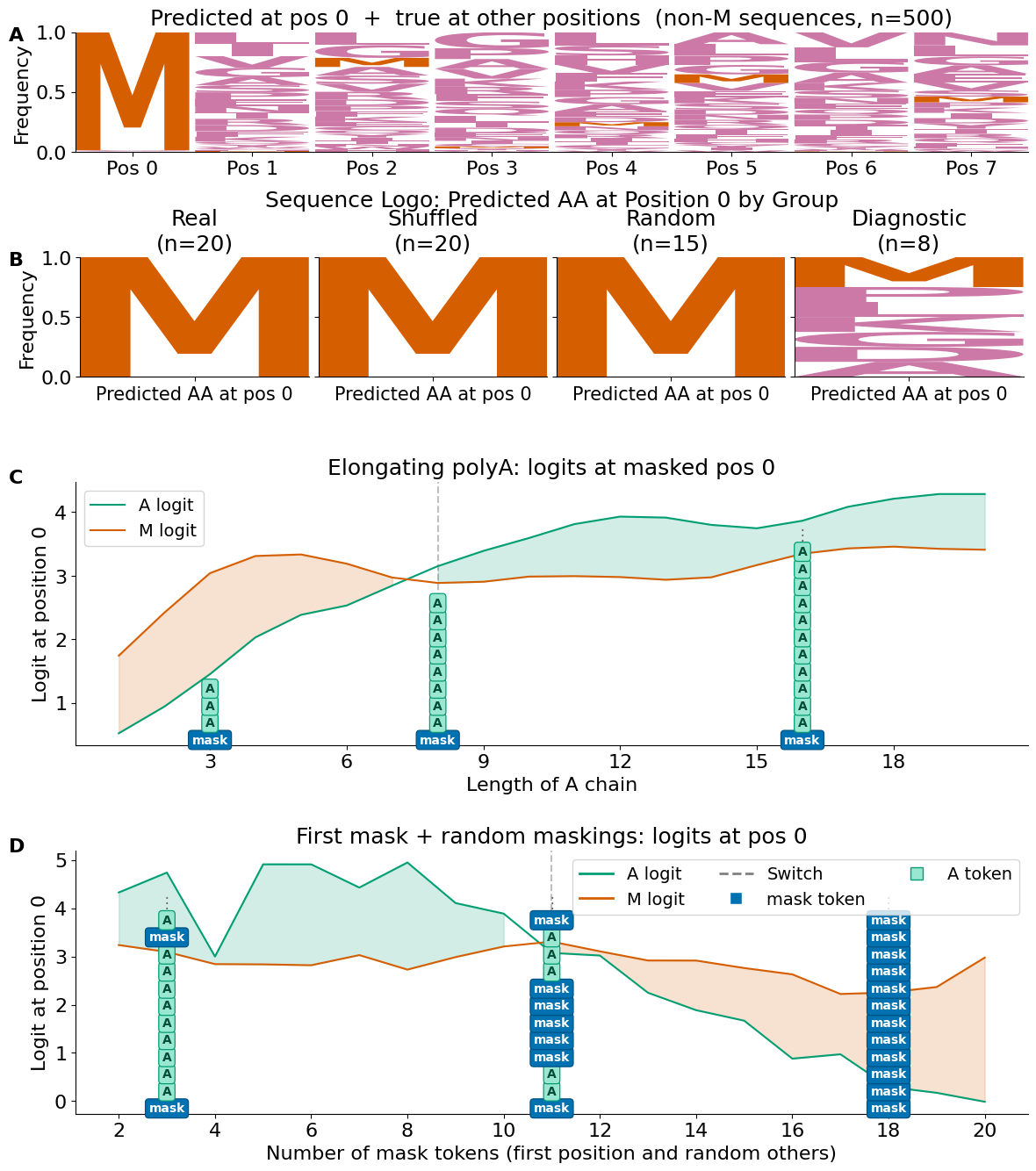}
    \caption{\textbf{Methionine prediction at position~0 is a robust positional
    prior that emerges from circuit competition.}
    \textbf{(A)}~Predicted amino acids (only pos0 is masked) for 500 UniProt sequences
    whose true first residue is not methionine. The model overwhelmingly
    predicts M confirming a positional
    prior that overrides sequence-specific evidence.
    \textbf{(B)}~Distribution of M's rank in the model's predictions when
    position~0 is masked across 63 sequences spanning four categories: real
    UniProt proteins, composition-matched shuffled controls, random amino acid
    strings and diagnostic sequences. The
    failures are exclusive to polyamino-acid sequences.
    \textbf{(C)}~Context elongation experiment. A MASK at position~0 is
    followed by poly-A chains of increasing length (1--20 residues).The A logit rises
    steadily while the M logit remains approximately constant, with a
    crossover at $\sim$9--10 residues (dashed line).
    \textbf{(D)}~Context destruction experiment. Starting from a
    20-residue Poly-A sequence with MASK at position~0, additional masks
    are introduced at random internal positions (averaged over 50 random
    orderings). The A logit decreases as sequence context is
    destroyed while the M logit remains stable; M re-emerges as the
    dominant prediction at $\sim$8--10 total masks.}
    \label{fig:robustness_competition}
\end{figure}

To quantify this effect, we masked the first residue in 500 UniProt sequences whose true N-terminal amino acid is not methionine (Fig. ~\ref{fig:robustness_competition} A). The model predicted methionine in the majority of cases, indicating a strong positional prior that overrides sequence-specific information. The same pattern held across a deliberately heterogeneous set of sequences including shuffled proteins, random amino-acid strings, and diagnostic constructs (Fig. 2 B; see Methods). Methionine was the top prediction in 57 of 63 cases, confirming that the prior is largely independent of sequence content.

The six failures were informative. All were poly-amino-acid sequences (e.g. poly-A, poly-G) in which the repeated residue took rank 1, and in every case methionine was ranked second. The prior is therefore not abolished by repetitive context — it is merely outvoted. This already suggests that prediction reflects competition between signals rather than conditional activation of a rule. Rather than switching methionine on at position 0 (first position), the model appears to combine multiple signals whose relative strength determines the output.

To test this hypothesis, we performed two complementary perturbations. In a context elongation experiment, a masked first position was followed by poly-alanine chains of increasing length. As the chain lengthened, the alanine logit, at the first position, rose steadily, while the methionine logit remained approximately constant (Figure~\ref{fig:robustness_competition} C and Extended Data ~\ref{fig:ext-polyA-experiments}). At roughly nine to ten residues, alanine overtook methionine as the top prediction. In a context destruction experiment, we progressively masked residues within a poly-alanine sequence. As contextual information was removed, the alanine logit decreased, while the methionine logit again remained stable (Fig.~\ref{fig:robustness_competition} D and Extended Data ~\ref{fig:ext-polyA-experiments}). Methionine re-emerged as the dominant prediction once sufficient context was destroyed.

These experiments reveal two properties. The methionine circuit contributes a roughly constant signal at the first position, and prediction changes only when competing circuits produce stronger alternatives. A causal test confirms this directly: injecting L6H8's activations from methionine-predicting sequences into poly-amino-acid targets — artificially strengthening the methionine circuit's output — fails to rescue methionine prediction in any target (Extended Data Fig. 16). Competition, not circuit strength, determines the outcome. The model resolves its output through circuit competition rather than explicit gating: a methionine-favouring signal is always present at the first position, but can be overridden when context-dependent circuits become sufficiently strong.

Because prediction depends on relative logit values rather than absolute magnitude, we use the M-gap — the methionine logit minus the strongest competing logit — as the primary diagnostic throughout the remainder of the analysis ( see Methods \nameref{sec:methods-mgap})). Competition between circuits explains the behavioral pattern, a constant methionine signal that loses to a sufficiently strong context. However, which components generate this constant signal? and how do those components reliably interact through so many different sequences? 

% --- FIGURE 2: Robustness + Competition ---

% -----------------------------------------------------------------------------
\subsection{The methionine circuit is composed of a minimal set of attention heads and MLPs}
\label{sec:circuit}

If methionine prediction arises from a specific circuit rather than distributed computation, knocking out (ablating) key components should disrupt the behaviour. We therefore zeroed each of the twelve attention and MLP modules at the masked position and measured the effect on methionine prediction. Only Layer 6 attention caused widespread failure, breaking methionine prediction in 91$\%$ of sequences (Fig. ~\ref{fig:ablation}~A). This localises the readout of the positional prior to a late attention layer.

Head-level ablation within Layer 6 identified a single dominant component (Figure~\ref{fig:ablation}~B). Removing head L6H8 produced the largest drop in methionine prediction and the highest failure rate across sequences. Inspection of its attention pattern revealed near-exclusive focus on the BOS token (Extended Data
Fig. ~\ref{fig:ext-l6-attention}), suggesting that this head reads information stored at BOS and writes a methionine-favouring signal into the masked position.

Targeted knockout confirmed this interpretation. Disrupting only the BOS-directed attention of L6H8 produced nearly the same effect as ablating the entire head (r = 1.0; Extended Data Fig. 8). L6H8's function is therefore mediated entirely through its attention to BOS. This immediately raises a problem. BOS is present for every masked token in every sequence, and L6H8 is present in every forward pass, yet the model predicts methionine almost exclusively at position 0. Something upstream must be telling L6H8 when to act.

\subsection{The circuit requires upstream query composition}

\begin{figure}[htbp!]
    \centering
    \includegraphics[width=\textwidth]{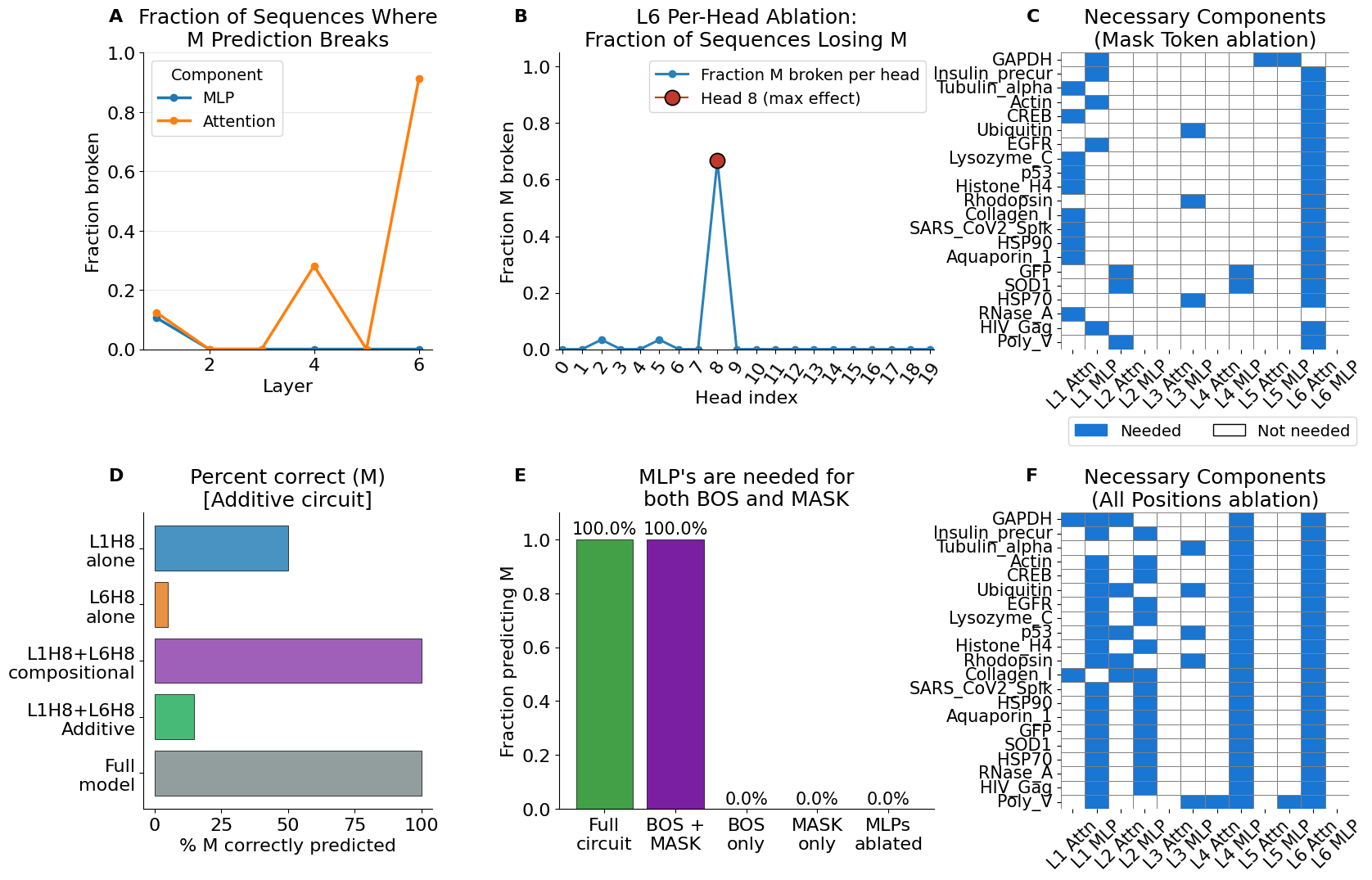}
    \caption{\textbf{Systematic ablation identifies a compact methionine
    circuit centred on L6H8 with upstream compositional support.}
    \textbf{(A)}~Fraction of sequences where M prediction breaks upon
    mask-only ablation (zeroing at the masked position only) of each
    layer attention and MLP modules. Only Layer~6 attention causes
    widespread failure (91\% of sequences).
    \textbf{(B)}~Head-level ablation within Layer~6: fraction of sequences
    losing M prediction per head. Head~8 (L6H8) ablation causes failure
    in $\sim$65\% of sequences; all other heads have a negligible impact.
    \textbf{(C)}~Per-sequence component necessity heatmap under mask-only
    ablation. Each row is a test sequence; coloured cells indicate that
    the component is required for M prediction. L6 attention and L1
    attention are the most frequently needed.
    \textbf{(D)}~Additive versus compositional (joint-head synergy test). M-prediction
    rate for five conditions: L1H8 alone (50\%), L6H8 alone (5\%), additive injection
    of both heads' outputs bypassing the computational graph (15\%),
    joint operation where L1H8's output feeds through the network into
    L6H8's query (100\%), and the clean baseline (100\%). The joint
    condition matches the baseline, confirming the query composition.
    \textbf{(E)}~MLP token-scope test within the minimal circuit.
    M-prediction rate for five conditions: full circuit with all core
    MLPs active (100\%), MLPs active only at BOS and MASK positions
    (100\%), MLPs active only at BOS (0\%), MLPs active only at MASK
    (0\%), and all core MLPs ablated (0\%). Preserving MLP processing
    at both BOS and MASK is necessary and sufficient; either alone
    fails.
    \textbf{(F)}~Per-sequence component necessity heatmap under
    all-positions ablation (zeroing at every token position). Under
    this stricter regime, early-layer MLPs (L1, L2, L4) emerge as
    critical alongside L6 attention, reflecting their role in building
    key/value representations at BOS and query representations at the
    masked position.}
    \label{fig:ablation}
\end{figure}

A greedy minimal-circuit search (see methods) identified two attention heads, L1H8 and L6H8, as the core components of the methionine circuit (Fig. ~\ref{fig:ablation}~A, ~B and C,
Table~\ref{tab:ext-greedy-heads}). The relationship between them was not additive. L6H8 alone rescued only a small fraction of predictions, and L1H8 alone produced partial rescue (Fig. ~\ref{fig:ablation}~D). Simply combining their outputs outside the computational graph yielded limited improvement (see methods). In contrast, allowing both heads to operate jointly restored methionine prediction across sequences.

This super-additive effect indicates query composition. L1H8 contributes to building the masked token representation that L6H8 later uses as its query. L6H8 does not act independently, but instead relies on upstream computation to determine when to attend to BOS. Methionine prediction therefore depends on a multi-layer interaction in which early heads shape the query consumed by a later readout head.

The ablation experiments so far preserved normal processing at all positions except the mask, including at BOS. When we extended ablation to every token position, a stricter test, the minimal circuit changed. Early-layer MLPs in Layers 1–4 now became necessary alongside Layer 6 attention (Fig. ~\ref{fig:ablation}E), with the most common minimal set given by \{L1 MLP, L2 MLP, L4 MLP, L6 Attention\}.

These MLPs build the representations that the circuit depends on. They shape the key and value vectors at BOS, and the query at the masked position, so that the inputs reaching Layer 6 match the expectations of L6H8. When these MLPs are removed, the representations deviate, and the circuit fails. This also explains why they were not required in the mask-only setting: there, BOS was still processed by the full model, providing L6H8 with intact key and value representations. Consistent with this, preserving only the relevant MLPs at BOS and the attention pathway at the masked position is sufficient to recover methionine prediction (Extended Data Fig. ~\ref{fig:ext-mlp-bos-attn-mask}).

A token-scope analysis sharpens this picture further. Preserving MLP processing at both BOS and the masked position is necessary and sufficient for methionine prediction, whereas preserving either alone fails (Fig. 3F).
Isolating this minimal subnetwork also produces the cleanest causal evidence for circuit competition. The isolated circuit predicts methionine at position 0 more strongly than the full model, but over-predicts methionine at internal positions (41\% vs. 0\% in the full model). This is a dissociation: the same positional circuit that is kept silent at internal positions in the full model fires freely when competing circuits are removed. Competition is therefore not inferred from correlational logit behaviour but demonstrated by manipulation — remove the competitors, and the prior is applied everywhere.

We now have the components of the circuit: early MLPs that build representations, upstream attention heads that compose a query, and L6H8 that reads from BOS and writes a methionine-specific signal. All of these components are present at every position. What differs between position 0 and internal positions is therefore not the presence of the circuit, but how it is engaged.

% --- FIGURE 3: Combined Ablation ---

% -----------------------------------------------------------------------------
\subsection{Positional specificity resides in the query}
\label{sec:causality}

This creates a puzzle. BOS is present in every sequence and L6H8 reads from it in every forward pass, yet the methionine signal only appears at position 0. Since the key and value representations at BOS are identical across runs, whatever distinguishes position 0 from an internal position must reside in the query formed at the masked token itself.

To test this, we performed activation patching between runs with the mask at position 0 and runs with the mask at internal positions. Patching BOS representations had little effect on methionine prediction. In contrast, patching the masked token representation at Layer 6 fully restored methionine prediction (Fig. ~\ref{fig:patching} A and B). This asymmetry indicates that BOS key and value representations are effectively position-invariant, whereas positional specificity is determined by the query.

Further analysis showed that both upstream-composed query content and the rotary positional embedding applied at Layer 6 were required. Neither component alone rescued prediction (Figure~\ref{fig:patching} C). The full post-RoPE query, combining both, restored methionine prediction in the majority of cases. Positional selectivity therefore arises from the interaction between learned query content and the geometry of the positional rotation.

Gradient analysis of the L6H8 output--value (OV) write vector (see methods) confirmed that this head functions as a methionine-specific writer.The cosine similarity between L6H8's OV vector and the gradient direction for the methionine logit was 0.35--0.4, substantially higher than any other Layer~6 head and any other amino acid direction (Fig. ~\ref{fig:patching} D). Together, these results establish that L6H8 reads a stable BOS reference using a position-specific query constructed upstream. L6H8 thus functions as a targeted ``Methionine-writer''.

\subsection{Upstream heads form a distributed positional encoder}

The dependence of L6H8 on upstream computation suggests that positional information is assembled gradually. A search over attention heads identified a small set spanning early and mid layers whose positional rotations contribute to methionine prediction. Swapping these heads’ RoPE rotations to position-0 values progressively restored methionine prediction at internal positions (Table~\ref{tab:ext-greedy-heads}).

\begin{figure}[htbp!]
    \centering
    \includegraphics[width=\textwidth]{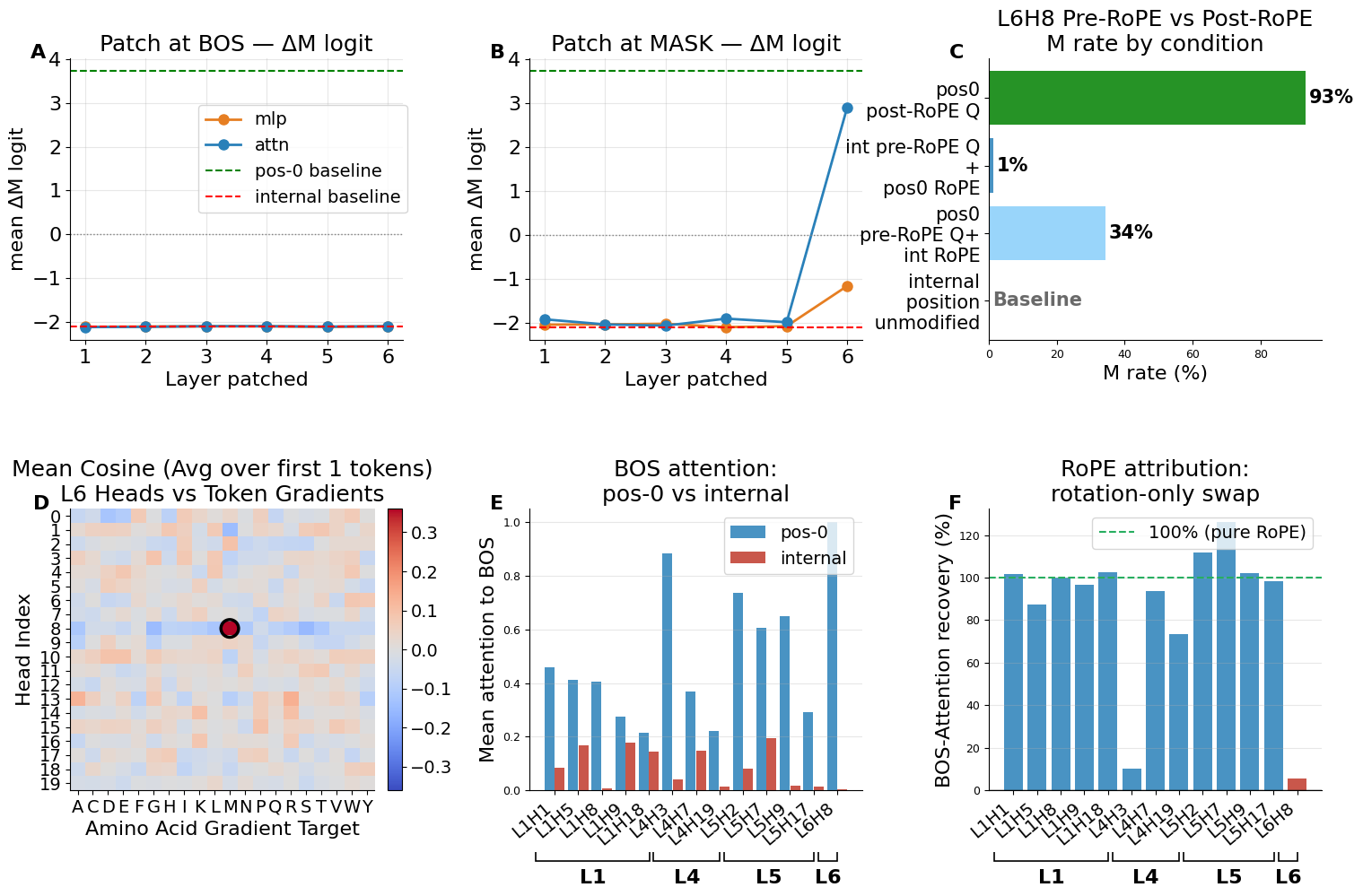}
    \caption{\textbf{Activation patching reveals query-mediated causality,
    methionine-specific output alignment, and a distributed upstream
    positional encoder.}
    \textbf{(A)}~M-gap when patching BOS activations from a position-0
    (``clean'') run into an internal-position (``corrupted'') run, layer
    by layer. Patching BOS has no effect: both MLP and attention traces
    remain near the corrupted baseline (red dashed line), indicating that
    BOS representations are effectively position-invariant.
    \textbf{(B)}~M-gap when patching MASK activations. Patching MASK
    attention output at Layer~6 fully rescues M prediction to the clean
    baseline (green dashed line), establishing that positional specificity
    resides entirely in the query formed at masked positions.
    \textbf{(C)}~Pre- versus post-RoPE query patching at L6H8. Neither
    position-0 pre-RoPE query content with internal-position RoPE
    (M-rate: 34\%) nor internal pre-RoPE content with position-0 RoPE
    (M-rate: 1\%) is sufficient alone; the full post-RoPE query
    (combining both) rescues 93\% of predictions, demonstrating synergy
    between upstream-composed content and RoPE-mediated geometric
    alignment.
    \textbf{(D)}~Heatmap of mean cosine similarity between all Layer~6
    heads' output--value (OV) write vectors and logit gradient directions
    for each amino acid. The L6H8~$\times$~M cell (circled) is distinctly
    highlighted ($\sim$0.37), confirming that L6H8 functions as a
    targeted methionine-specific writer.
    \textbf{(E)}~Mean attention weight to BOS at position~0 (blue) versus
    internal positions (red) for all key upstream heads and the L6H8
    readout head. Most upstream heads show substantially higher BOS
    attention when the MASK is at position 0.
    \textbf{(F)}~RoPE attribution: fraction of each head's BOS attention
    preference recovered by swapping RoPE rotations alone
    ($\sim$100\% = purely positional mechanism). Most upstream heads are
    purely RoPE-driven; L6H8 and L4H3 are
    context-dependent exceptions, confirming that L6H8 requires the
    correctly composed query rather than relying on positional signals
    alone.}
    \label{fig:patching}
\end{figure}

Most of these heads preferentially attend to BOS when the mask is at position 0. This preference is largely RoPE-driven, indicating that positional encoding is distributed across multiple heads (Fig.~\ref{fig:patching}E,F). Two exceptions stand out. L6H8 itself is strongly context-dependent, consistent with its role as a readout head. L4H3 is also largely context-dependent (RoPE alone explains only 10\%), suggesting it routes attention based on residual-stream content in addition to positional signals. Together, these results indicate that upstream heads encode position into the query, while L6H8 performs the final readout.

Necessity and sufficiency tests on the BOS pathway further clarify this division of labour (Table~\ref{tab:vbos}). Ablating BOS-derived value vectors across all non-readout heads caused only a modest reduction in methionine prediction (M-rate: 1.0~$\to$~0.825), whereas ablating L6H8’s BOS values was catastrophic (M-rate~$\to$~0.018).

This asymmetry identifies the BOS pathway as the dominant information channel. Upstream heads contribute redundantly to constructing the query, but the final readout depends critically on BOS-derived values at L6H8. Consistent with this, restricting the circuit to BOS-derived values alone preserves full methionine prediction. The upstream heads therefore act collectively to encode first-position identity into the query, largely through independent RoPE-driven effects. This distributed encoding ensures robust query formation, while concentrating specificity in a single readout step. This raises a final mechanistic question: what is RoPE geometrically doing that makes position 0 special?

% --- FIGURE 4: Activation Patching + Upstream ---

% --- TABLE 2: V_BOS necessity/sufficiency ---
\begin{table}[htbp!]
    \centering
    \small
    \caption{Necessity and sufficiency of BOS-derived value
    vectors. M-rate under ablation (necessity, position-0 setting)
    and RoPE-swap rescue (sufficiency, internal-position setting)
    for different ablation scopes.}
    \label{tab:vbos}
    \begin{tabular}{lcccc}
        \toprule
        Condition & \multicolumn{2}{c}{Necessity (pos-0)} &
        \multicolumn{2}{c}{Sufficiency (internal)} \\
        \cmidrule(lr){2-3} \cmidrule(lr){4-5}
        & M-rate & M-logit drop & M-rate & Rescue (\%) \\
        \midrule
        Baseline & 1.000 & --- & 0.793 & 55.6 \\
        \midrule
        \multicolumn{5}{l}{\emph{Non-readout core heads (excl.\ L6H8)}} \\
        \quad Ablate V$_\text{BOS}$ & 0.825 & +1.54 & 0.018 & 6.0 \\
        \quad Ablate V$_\text{ALL}$ & 0.474 & +3.69 & 0.007 & 3.9 \\
        \quad Keep only V$_\text{BOS}$ & 1.000 & +0.53 & 0.688 & 45.2 \\
        \midrule
        \multicolumn{5}{l}{\emph{All core heads (incl.\ L6H8)}} \\
        \quad Ablate V$_\text{BOS}$ & 0.018 & +5.48 & 0.000 & 2.9 \\
        \quad Ablate V$_\text{ALL}$ & 0.000 & +5.37 & 0.000 & 3.1 \\
        \quad Keep only V$_\text{BOS}$ & 1.000 & +0.53 & 0.698 & 46.5 \\
        \bottomrule
    \end{tabular}
\end{table}
\clearpage
% -----------------------------------------------------------------------------
\subsection{RoPE geometry gates the circuit}
\label{sec:rope}

Activation patching established that L6H8’s query depends on both upstream-composed content and the correct RoPE rotation. These are qualitatively different sources of information: one is learned through training, the other imposed by the architecture. How do they interact, and which drives the high BOS attention? To address this, we decompose the attention score as:

\begin{equation}
    \mathbf{Q} \cdot \mathbf{K}
    = \|\mathbf{Q}\| \, \|\mathbf{K}\| \cos\bigl(\theta_{\mathbf{Q},\mathbf{K}}\bigr),
    \label{eq:qk_decomp}
\end{equation}
where $\|\mathbf{Q}\|$ and $\|\mathbf{K}\|$ are the Euclidean norms
of the query and key vectors and $\theta_{\mathbf{Q},\mathbf{K}}$ is
the angle between them. This decomposition separates the attention score into two components: a magnitude term, determined by the norms of the query and key vectors, and an alignment term, determined by their angular relationship. Under RoPE, query and key vectors are rotated in a frequency-dependent manner, making their alignment a direct function of relative position.

This framing allows us to ask a precise question: when L6H8’s attention to BOS decreases at internal positions, is this due to reduced query magnitude (a size effect), loss of angular alignment (a geometry effect), or both? If RoPE provides positional specificity, removing it should disrupt methionine prediction at position~0. It does not. Removing RoPE at the masked token in any single layer, including Layer~6, leaves prediction intact, although the M logit is reduced (Extended Data Fig. ~\ref{fig:ext-basic-rope-ablation}). This apparent contradiction with the patching results reflects an important distinction. Ablating RoPE removes a helpful signal, whereas applying the \emph{wrong} rotation (corresponding to an internal position) actively introduces angular misalignment that degrades BOS attention.

This distinction becomes clearer in the isolated minimal circuit. Removing RoPE does not impair methionine prediction at position~0, but instead causes the circuit to predict methionine \emph{everywhere} (Fig. ~\ref{fig:rope}B). RoPE is therefore not required to enable position-0 prediction, but to suppress prediction at internal positions.

In the full model, this suppression is handled by competing circuits. In the isolated circuit, RoPE is the only mechanism that can distinguish position~0 from internal positions. Thus, RoPE acts not only as a positional signal, but as a gate that prevents the methionine prior from being applied everywhere. This reframes the problem: if RoPE both enables and suppresses the circuit, what determines where it is active?

\subsection{Positional selectivity arises from coupled norm and alignment changes across frequency bands}
\label{sec:rope-freq}

Futher, we analyse the contribution of individual RoPE frequency bands. Under RoPE, the attention score decomposes additively across frequency pairs ~\citep{Su2023RoFormer:Embedding} :

\begin{equation}
    \mathbf{Q}_{\text{MASK}} \cdot \mathbf{K}_{\text{BOS}}
    = \sum_{f=0}^{n_f - 1} \mathbf{Q}^{(f)}_{\text{MASK}}
      \cdot \mathbf{K}^{(f)}_{\text{BOS}},
    \label{eq:rope_decomp_sum}
\end{equation}
where $\mathbf{Q}^{(f)}$ and $\mathbf{K}^{(f)}$ denote the
two-dimensional sub-vectors associated with frequency pair $f$.
Within each band we can further factor the dot product as:
\begin{equation}
    \mathbf{Q}^{(f)}_{\text{MASK}} \cdot \mathbf{K}^{(f)}_{\text{BOS}}
    = \bigl\|\mathbf{Q}^{(f)}_{\text{MASK}}\bigr\| \,
      \bigl\|\mathbf{K}^{(f)}_{\text{BOS}}\bigr\| \,
      \cos\!\bigl(\theta^{(f)}_{\mathbf{Q},\mathbf{K}}\bigr),
    \label{eq:per_freq_decomp}
\end{equation}
where $\theta^{(f)}_{\mathbf{Q},\mathbf{K}}$ is the angle between the
query and key subvectors in band $f$ after RoPE rotation.

To examine this, we decompose the score difference between position~0 and internal positions into per-frequency contributions, separating changes in query magnitude from changes in angular alignment. This decomposition has a key property. Because the RoPE rotation matrix $\mathbf{R}^d_{\Theta,m}$ is orthogonal~\citep{Su2023RoFormer:Embedding}, it cannot alter vector norms. Hence any difference in $\|\mathbf{Q}^{(f)}\|$ between position~0 and internal positions must therefore originate entirely in the \emph{pre-RoPE} residual stream—that is, in the query content written by upstream attention heads. In contrast, $\|\mathbf{K}^{(f)}_{\text{BOS}}\|$ is determined by BOS processing and is effectively position-invariant (\nameref{sec:causality}). This separation allows us to isolate the role of upstream query composition. Does it increase L6H8’s attention to BOS by amplifying the query (a size effect), or by rotating it into alignment with the BOS key (a geometric effect), or both? These alternatives correspond to distinct mechanisms: amplification scales an existing signal, whereas alignment changes which key the query matches. To our knowledge, this norm–cosine decomposition has not previously been applied as an interpretability tool in transformer models. Rather than decomposing the parts, previous studies have largely treated the attention scores as fixed values ~\citep{NelsonElhage2021ACircuits, Marks2025SparseModels}.

Each band contributes $|\mathbf{Q}^{(f)}| \cdot |\mathbf{K}^{(f)}| \cdot \cos(\theta^{(f)})$ to the attention score, and this product behaves as an AND gate: norm sets the gain, cosine sets the alignment, and both must be non-negligible for the band to register. Because RoPE is orthogonal it cannot alter sub-vector norms (see Methods), and $|\mathbf{K}^{(f)}_{\text{BOS}}|$ is effectively position-invariant (Section~\ref{sec:causality}; Extended Data Fig.~\ref{fig:rope-freq-dec}C,F), positional asymmetry must enter through the query representation—either through its angle relative to the BOS key, its per-band norm, or both. Non-BOS key suppression can also be ruled out: the activation patching in Section~\ref{sec:causality} shows L6H8's positional selectivity is fully explained by the query.

The two query-side factors play complementary roles. Cosine alignment is what selects BOS as the argmax: within a single query position, $|\mathbf{Q}^{(f)}|$ multiplies all query–key scores equally, so it cannot, on its own, distinguish BOS from other keys. Selection must therefore arise from cosine alignment. Per-band norm, in contrast, scales how much a given band contributes to every score; because the softmax is sensitive to score magnitude, a larger contribution from any one band sharpens the resulting attention distribution onto whichever key is selected. Cosine determines the ranking of keys; norm controls the contrast between them. Mechanisms resembling learned gain or temperature control in attention have been proposed as ways to regulate attention sharpness. Here, a similar effect emerges endogenously: upstream RoPE-mediated heads construct a position-dependent query whose norm asymmetry modulates attention sharpness at L6H8.

Different bands take different routes to position-0 selectivity (Extended Data Fig.~\ref{fig:rope-freq-dec}I). The dominant band $f_0$ (period $\approx 6.28$ tokens) operates on both axes: $\|\mathbf{Q}^{(f_0)}\|$ falls roughly seven-fold from position~0 to an internal position (1.14 vs.\ 0.15 at position~10; Extended Data Fig.~\ref{fig:rope-freq-dec}G), and cosine alignment with BOS drops from $\sim$0.8 to near zero (Extended Data Fig.~\ref{fig:rope-freq-dec}H). The symmetric attribution accordingly gives $f_0$ a balanced 50/50 norm--direction split, with the largest reference-choice spread of any band because both factors change substantially. That $f_0$ also drives the cycling of M prediction under RoPE shifts (Figure~\ref{fig:rope}A,C) confirms that this band couples directly to the model's output: shifting Q-RoPE rotates $\mathbf{Q}^{(f_0)}$ relative to $\mathbf{K}^{(f_0)}$, producing the $\sim$6-token cosine oscillation seen in Figure~\ref{fig:rope}C, and the $f_0$ norm sets the amplitude of the resulting effect on the M logit.

The secondary bands ($f_1$, $f_2$, $f_6$) act primarily as alignment selectors: their cosines with BOS collapse --- $f_1$, $f_2$, and $f_6$ all flipping sign --- at internal positions (Figure~\ref{fig:rope}F,G; Extended Data Fig.~\ref{fig:rope-freq-dec}E,H), while norms change little. Attribution in these bands is 80--98\% direction-driven.

At internal positions, both routes fail in tandem: cosine alignment collapses across $f_0$, $f_1$, $f_2$, and $f_6$, removing BOS as the argmax target across multiple bands, and the $f_0$ norm drops seven-fold, removing the gain that would otherwise sharpen attention onto the selected key (Extended Data Fig.~\ref{fig:rope-freq-dec}D). Globally, 39\% of the positional score difference is norm-attributable and 61\% direction-attributable (Table~\ref{tab:ext-rope-decomp}). Both mechanisms are real and necessary: the AND gate closes simultaneously on alignment and gain.

\subsection{RoPE shifting reveals periodic positional geometry}
\label{sec:rope-shift}

\begin{figure}[htbp!]
    \centering
    \includegraphics[width=1.1\textwidth]{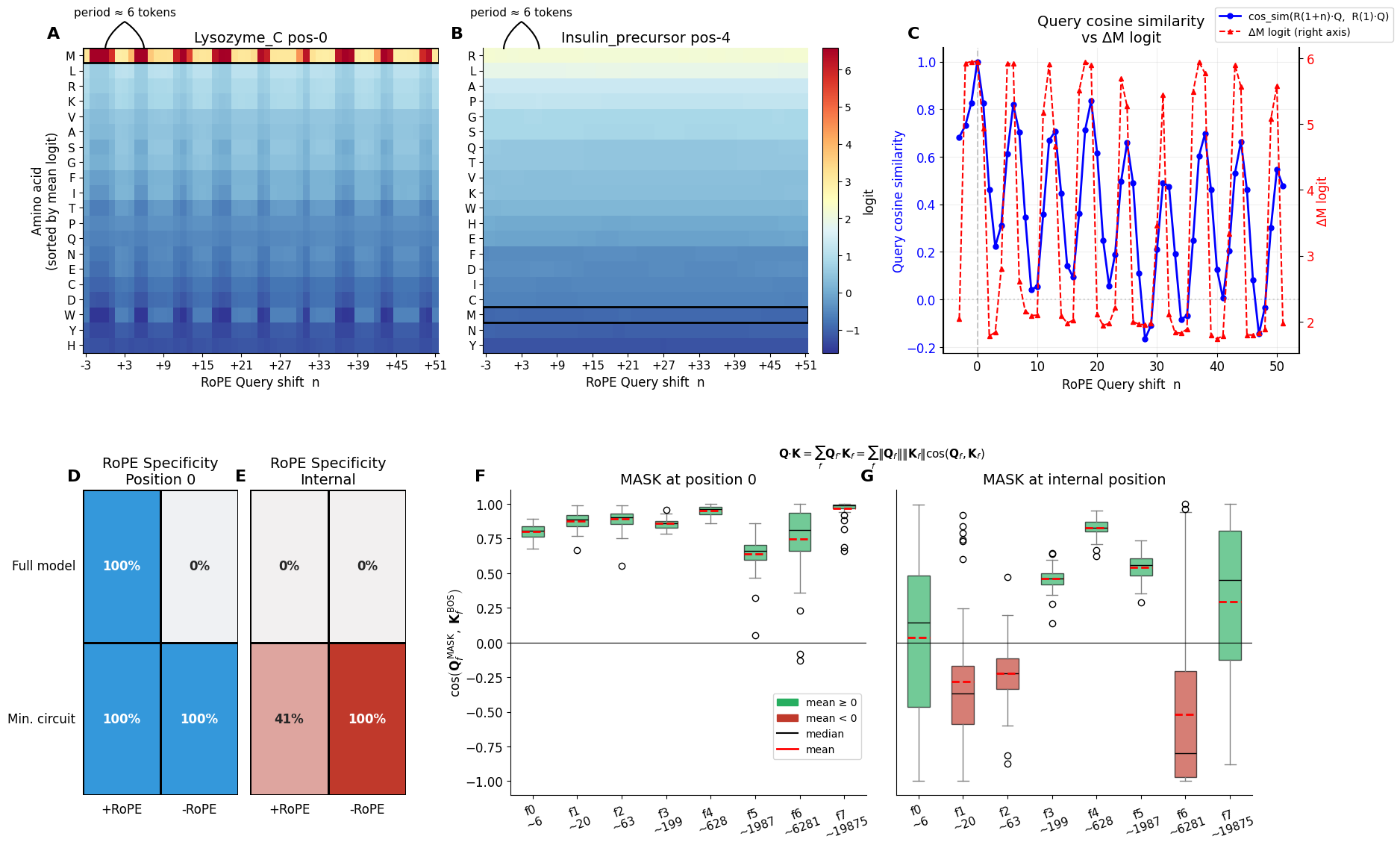}
    \caption{\textbf{Rotary position embeddings provide the geometric basis
    for positional selectivity.}
    \textbf{(A)}~Full logit landscape under query-side RoPE shifts for
    Lysozyme with MASK at position~0. Amino acid logits (rows, sorted by
    mean) are shown as a function of RoPE shift $n$ ($-3$ to $+50$).
    Methionine (top row) displays clear cyclic dominance with
    $\sim$6-token periodicity, matching the dominant RoPE frequency $f_0$.
    \textbf{(B)}~Same logit landscape as (A) but for Insulin with MASK at
    an internal position (Position 4). The clean periodicity is disrupted;
    competing amino acids (R, L, A) dominate and methionine prediction
    remained suppressed across all shifts.
    \textbf{(C)}~Query cosine similarity (blue, left axis) versus M-gap
    (red dashed, right axis) as a function of Q-RoPE shift $n$ for
    Lysozyme. The two traces co-vary tightly ($\sim$6-token period),
    confirming that angular alignment between L6H8's query and BOS key
    vectors drives methionine prediction.
    \textbf{(D, E)}~Effect of RoPE on M-prediction specificity
    (2$\times$2 design). M-prediction rate at position~0 (blue) \textbf{(D)}~versus
    internal positions (red) \textbf{(E)}~for four conditions: full model with RoPE
    (100\% / 0\%), full model without RoPE (100\% / 0\%), minimal circuit
    with RoPE (100\% / 41\%), and minimal circuit without RoPE
    (100\% / 100\%). In the full model, competing circuits suppress
    internal M prediction regardless of RoPE; in the isolated circuit,
    RoPE is required to prevent ubiquitous M prediction.
    \textbf{(F)}~RoPE frequency decomposition with MASK at position~0
    ($n = 57$ M-predicting sequences). Cosine similarity between query
    and key components per frequency pair; values are uniformly high,
    confirming strong angular alignment across all bands.
    \textbf{(G)}~Same decomposition with MASK at an internal position (position~10). Cosine similarities drop sharply across multiple bands, with $f_1$, $f_2$ and $f_6$ flipping sign and $f_0$ collapsing from $\sim 0.8$ to near zero --- removing BOS as the alignment target across the bands that produced the largest position-0 scores. BOS key-vector norms remain effectively position-invariant (not shown; see Extended Data Fig.~\ref{fig:rope-freq-dec}C,F). The full positional difference between position~0 and internal positions reflects the loss of angular alignment in $f_1, f_2, f_6$ and a seven-fold collapse of the $f_0$ query norm; in $f_0$, both factors change substantially and contribute roughly equally to the score difference (Extended Data Fig.~\ref{fig:rope-freq-dec}G,I).}
    \label{fig:rope}
\end{figure}

To establish a direct causal link between RoPE geometry and methionine prediction, we applied a uniform index shift to L6H8’s query-side rotations: $R(i) \to R(i+n)$, leaving key-side rotations unchanged. As $n$ varied from $-3$ to $+50$, the M-gap oscillated with a period of approximately 6 tokens (Figure~\ref{fig:rope}A,C), matching the dominant frequency $f_0$ (period $\approx 6.28$ tokens; Extended Data
Table~\ref{tab:ext-rope-freq}). The cosine similarity between the original position-0 query and each shifted query closely tracked this oscillation (Fig. ~\ref{fig:rope}B), confirming that the effect arises from RoPE-mediated query–key geometry.

At internal positions, this periodicity is disrupted by competing circuits (Fig. ~\ref{fig:rope} C). However, shifting the RoPE rotation to mimic position~0 restores methionine prediction (Extended Data
Fig.~\ref{fig:ext-rope-shift}), providing direct causal evidence that positional selectivity is geometrically encoded.

% --- FIGURE 5: RoPE ---

\clearpage
% =============================================================================
% DISCUSSION
% =============================================================================
\section{Discussion}

We have traced the mechanistic basis of start-codon prediction in ESM2-8M from a behavioural observation to a circuit-level account. The model does not detect methionine locally at the masked position. Instead, it retrieves a methionine-favouring signal from a stable reference representation at the BOS token, using a position-specific query constructed across multiple layers via rotary positional embeddings. This signal is written by a single attention head (L6H8), whose output is aligned with the methionine logit. Specificity—correctly refraining from predicting methionine at internal positions—does not arise from inhibition within this circuit, but from competition with context-dependent circuits that produce stronger alternatives.

\paragraph{What this reveals about how PLMs use information.}
The most immediate implication is that this prediction depends almost entirely on positional information rather than sequence context. The model has internalised a simplified translation rule—that proteins begin with methionine—but not the more nuanced biology of N-terminal methionine excision. This illustrates how a model can appear to ``know'' biology while encoding a simpler statistical regularity, and highlights the value of mechanistic analysis in distinguishing between the two.

More broadly, the competition framework provides a lens for interpreting PLM predictions. Model outputs reflect the superposition of multiple circuits, each contributing its own signal. A high-confidence prediction does not necessarily indicate that a single decisive feature has been identified; it may instead reflect one circuit dominating others. Understanding which circuits contribute, and what information they use, is therefore essential for assessing whether a prediction rests on biologically meaningful evidence.

Our analysis focuses on ESM2-8M, a small model with six layers and eight million parameters. Whether the same circuit architecture persists in larger models or across PLM families remains an open question. However, the motifs we identify—query composition and the use of BOS as a reference node—have clear parallels in natural language models~\citep{Olsson2022In-contextHeads, Ruscio2025WhatSink}, suggesting they may reflect general computational strategies rather than model-specific quirks.

\paragraph{BOS as an active computational node.}
A central and unexpected finding is the role of the BOS token. In encoder-only PLMs, BOS is typically treated as a passive delimiter and often discarded in downstream applications~\citep{Vieira2024Medium-sizedDatasets}. Our results show that it instead functions as an active computational node, storing a position-invariant key and value representation that the circuit retrieves from. Its fixed position provides a geometric anchor for RoPE-based positional selectivity, allowing queries to target it reliably across sequences.

This interpretation aligns with the reference-frame hypothesis of~\citet{Ruscio2025WhatSink}, which proposes that attention-sink tokens act as geometric anchors in transformer representations. Our results extend this idea to encoder protein language models with RoPE and provide a causal mechanism through which such a reference frame is used. This contrasts with accounts in which attention to the first token serves primarily to prevent representational collapse~\citep{Barbero2025WhyToken}. In the methionine circuit, BOS-directed attention is not a passive sink but an active retrieval mechanism.

These observations suggest that discarding BOS representations may remove information the model actively uses. More generally, special tokens in transformer models may provide stable reference points that circuits exploit, a possibility worth exploring in future work.

\paragraph{Circuit competition as a general principle.}
The competition we observe suggests a general principle for transformer computation. The methionine circuit contributes a roughly constant signal; what varies across positions is the strength of competing circuits. This allows a simple positional prior to coexist with arbitrary context-dependent signals without requiring explicit suppression mechanisms. It also explains the redundancy we observe upstream: multiple heads contribute small, largely independent signals that ensure the positional circuit is strong enough to dominate under typical conditions.

For practitioners, this implies that model predictions are best understood as the outcome of competing signals rather than a single decision. Examining the distribution of logits, rather than only the top prediction, may therefore be informative: closely spaced logits suggest genuine competition between circuits, whereas a dominant logit indicates that one circuit acts largely unopposed.

\paragraph{Positional encoding operates through dual mechanisms.}
Positional information in RoPE-based transformers is not added explicitly but applied through rotations at each layer. The methionine circuit exploits this through a distributed set of upstream heads that encode position into the query, which L6H8 then uses to retrieve BOS information. The synergy between learned query content and RoPE rotation is essential: neither alone is sufficient, but together they recover the majority of predictions.

The frequency decomposition reveals that this synergy arises from coupled changes in query norm and angular alignment, distributed across frequency bands. The dominant short-period band ($f_0$) engages both factors: at internal positions, its query norm collapses by roughly seven-fold and its cosine alignment with the BOS key drops from $\sim 0.8$ to near zero. Secondary bands ($f_1, f_2, f_6$) contribute primarily through angular alignment, with cosines flipping sign at internal positions while norms change little. This combination allows the model to both amplify and geometrically target the BOS representation, and is more flexible than a single positional encoding mechanism. While prior work has emphasised angular misalignment in high-frequency bands \citep{Barbero2025RoundUseful}, our results show that norm-based amplification can play a comparably important role and that the two factors can co-vary within a single band.

\paragraph{Relation to known circuit motifs and discovery methods.}
The multi-layer query composition we identify is reminiscent of Q-composition in induction heads in language models~\citep{Olsson2022In-contextHeads}, where early layers write information that later layers use to form queries. Here, however, the composition is driven primarily by positional signals rather than token identity, representing a distinct functional motif. The upstream heads, which attend preferentially to BOS via RoPE-driven effects, resemble positional heads identified in large language models~\citep{Barbero2025RoundUseful}. The recurrence of these motifs across domains suggests that transformers converge on similar strategies for routing positional information.

\paragraph{Future directions.}
Several open questions follow from this work. The most immediate is identifying the competing circuits that suppress methionine at internal positions. Completing this picture would clarify how positional and contextual information are integrated more broadly. The heterogeneity of upstream heads is itself informative. Most heads in the composition circuit are purely RoPE-driven, but L4H3 is not — its BOS attention depends on residual-stream content that we have not yet identified. This suggests that upstream query composition can be gated by either architectural position (RoPE) or learned content features, and that these two gating modes coexist within a single circuit. Identifying what L4H3 reads, and whether content-gated upstream heads are a general feature of PLM circuits or specific to this case, is a natural next step.

Extending this analysis to predictions with richer biological content is the critical next step. Conservation signals, secondary structure propensities, and functional motif recognition all involve more complex interactions between position and context than methionine prediction, and each offers a concrete target for the methodology developed here. The norm--direction decomposition of RoPE frequency bands, in particular, could distinguish whether a prediction relies on retrieval of a positional default (with characteristic norm and alignment signatures at specific indices) or on recognition of an evolutionary or structural pattern (with alignment signatures driven by sequence content rather than position). Applied to variant-effect predictions, where the clinical stakes of the retrieval-versus-recognition distinction are highest, this approach would test whether model confidence reflects biological evidence or a statistical default — a distinction that confidence scores alone cannot provide.

% =============================================================================
% METHODS
% =============================================================================
\section{Methods}

\subsection{Model and sequence data}
\label{sec:methods-data}

All experiments used the 6-layer, 8-million-parameter ESM2 model
(\texttt{esm2\_t6\_8M\_UR50D})~\citep{Lin2023Evolutionary-scaleModel}. The evaluation dataset comprised 63
sequences across four categories: 20 real UniProt/SwissProt proteins (for example,
Insulin, Lysozyme, Ubiquitin, Actin), 20 composition-matched shuffled controls, 15
uniformly random amino acid strings (15--45 residues), and 8 diagnostic sequences
(poly amino-acid chains and short or M-rich sequences). For all tests, position~0
was replaced with \texttt{<mask>} token. Non-M-starting sequence validation used 500 UniProt sequences whose native first residue is not methionine.

\subsection{Ablation framework}

A custom PyTorch hook-based framework enabled zero-ablation (setting outputs to
zero) of specific attention or MLP modules at the targeted layers. Two scopes were used:

\emph{Mask-only ablation.} Component outputs were zeroed exclusively at the
\texttt{<mask>} position, preserving normal processing elsewhere. This isolates the local dependencies at the prediction site.

\emph{All-positions ablation.} Component outputs were zeroed at every token
position, including BOS and EOS. This reveals globally necessary modules.

For head-level ablation within a layer, individual heads were zeroed by nullifying
their contribution to the attention output prior to the output projection.

\subsection{M-gap as the primary diagnostic}
\label{sec:methods-mgap}

The softmax function that converts logits to probabilities is a smooth
approximation to argmax~\citep{Goodfellow2016DeepLearning}: the model's
prediction is determined not by the absolute value of any single logit but by its
rank among all tokens. The competition experiments
(Figure~\ref{fig:robustness_competition}~C) illustrate this directly: the
methionine logit remains approximately constant while the alanine logit rises,
and the prediction flips precisely at the crossover point. We therefore define
the M-gap as the methionine logit minus the maximum competing logit. This
quantity directly determines whether methionine is the top-ranked prediction, and
is the one whose causal structure the remainder of the paper seeks to explain.

\subsection{Greedy minimal-circuit discovery}

Starting from the unablated model, we iteratively identified the least-impactful
component (by M-logit drop) and permanently ablated it. This continued until the methionine prediction failed, yielding the minimal necessary set. The procedure
was repeated for each test sequence, and component necessity frequencies were
aggregated. Both mask-only and all-position scopes were tested.

\subsection{Causal sufficiency testing}

We performed forward passes with all components zero-ablated except a specified
subset. This inverts necessity-based ablation: rather than removing one component,
we remove everything except the hypothesised circuit and test whether it reproduces
the target behaviour. This approach was used to test the L1+L6 attention subcircuit, MLP-supplemented subcircuit, and full minimal circuit.

\subsection{Activation patching}

Activation patching was performed between ``clean'' (MASK at position~0) and
``corrupted'' (MASK at an internal position) forward passes of the same sequence.
During the clean pass, intermediate representations (layer inputs, attention
outputs, query/key/value vectors) were cached. These were selectively injected into the corrupted pass via forward pre-hooks. We patched at the granularity of full
layer outputs, individual head outputs, specific token positions (MASK or BOS), and
specific attention sub-components (Q, K, V vectors and pre- versus post-RoPE
representations).

\subsection{Cross-sequence activation injection}

To test whether L6H8's output can override competing signals, we extracted L6H8's
pre-projection activations at the MASK token from ``source'' sequences
(M-predicting) and injected them into ``target'' sequences (poly-amino-acid chains
where M prediction fails). The M-gap was tracked before and after the injection.

\subsection{RoPE manipulation}

\emph{RoPE ablation.} RoPE rotations were removed (set to identity) at specified layers and token positions during forward pass.

\emph{RoPE swapping.} During internal-position-MASK runs, the native RoPE rotations at the specified heads were replaced with position-0 rotations. A greedy search identified the minimal head set for which this swap rescued the M prediction.

\emph{Frequency decomposition.} The pre-softmax score
$\mathbf{Q}_\text{MASK} \cdot \mathbf{K}_\text{BOS}$ was decomposed into
contributions from each of 8 RoPE frequency pairs (head dimension~$= 16$,
non-interleaved format). Per-pair dot products, cosine similarities, and vector
norms were computed across all M-predicting sequences.

\emph{Global index shift.} L6H8's query-side RoPE was uniformly shifted:
$R(i) \to R(i+n)$ for shift parameter $n \in [-3, +50]$, while key-side rotations remained at native indices. A control shifting both Q and K by the same offset
confirmed a residual difference $< 0.00002$.

\subsection{Logit gradient analysis}

The gradient of the methionine logit with respect to the final residual stream (post-Layer~6, pre-LayerNorm) was computed at the masked position. Each Layer~6 head's OV write vector ($\sum_j \alpha_{t,j} \mathbf{W}_O \mathbf{v}_j$) was extracted, and cosine similarities with all 20 amino acid logit gradients were computed and averaged across the test set.

\subsection{Competition experiments}

\emph{Context elongation.} Sequences were constructed with a MASK at position~0
followed by poly-A chains of length 1--20. The full logit vector and softmax probabilities were recorded at the masked position for each length.

\emph{Context destruction.} Starting from a 20-residue Poly-A sequence with MASK at
position~0, additional masks were introduced at random internal positions (2--20
total masks), averaging over 50 random masking orders per condition.

\subsection{Evaluation metrics}

Primary metrics were: M-gap (methionine logit minus maximum competing logit),
M-rate (fraction of sequences with M as top-1 prediction), P(M) (softmax
probability of methionine), M-logit drop (decrease in raw M logit upon
intervention), rescue percentage ((patched M-gap $-$ corrupted M-gap) / (clean
M-gap $-$ corrupted M-gap) $\times$ 100), and rescue rate for cross-sequence
injection (fraction of targets where M becomes top-1).

\subsection{Symmetric norm--direction decomposition of RoPE frequency bands}
\label{sec:rope-decomposition}

To quantify the relative contributions of query-norm amplification and
angular realignment to L6H8's positional score difference, we decomposed
$\Delta S_f = S^{(f)}_{\text{pos0}} - S^{(f)}_{\text{int}}$ for each
RoPE frequency band $f$, where
$S^{(f)} = \mathbf{Q}^{(f)}_{\text{MASK}} \cdot \mathbf{K}^{(f)}_{\text{BOS}}$
is the per-band dot product contributing additively to the pre-softmax
attention score (Equation~\ref{eq:rope_decomp_sum}).

\paragraph{Key reference.}
Because the BOS key representation showed a small 2--7\% positional
variation in norm across runs, we used the mean key sub-vector
$\mathbf{K}^{(f)}_{\text{avg}} = \tfrac{1}{2}(\mathbf{K}^{(f)}_{\text{pos0}}
+ \mathbf{K}^{(f)}_{\text{int}})$ as a neutral reference rather than privileging either condition. The score difference is then expressed
exactly as:
\begin{equation}
    \Delta S_f
    = \Delta\mathbf{Q}^{(f)} \cdot \mathbf{K}^{(f)}_{\text{avg}},
    \quad
    \Delta\mathbf{Q}^{(f)} = \mathbf{Q}^{(f)}_{\text{pos0}}
                             - \mathbf{Q}^{(f)}_{\text{int}}.
    \label{eq:delta_S}
\end{equation}

\paragraph{Symmetric linear attribution.}
Given $\Delta\mathbf{Q}^{(f)}$, we decomposed the score change into a
component attributable to the change in query-vector norm (the
\emph{norm component}) and a component attributable to the change in query direction (the \emph{direction component}). For a given reference
direction $\hat{\mathbf{q}}_{\text{ref}}$, the two components are
defined as:
\begin{align}
    \Delta\mathbf{Q}^{(f)}_{\text{norm}}
        &= \bigl(\|\mathbf{Q}^{(f)}_{\text{other}}\|
               - \|\mathbf{Q}^{(f)}_{\text{ref}}\|\bigr)
           \,\hat{\mathbf{q}}_{\text{ref}}, \\
    \Delta\mathbf{Q}^{(f)}_{\text{dir}}
        &= \Delta\mathbf{Q}^{(f)} - \Delta\mathbf{Q}^{(f)}_{\text{norm}},
\end{align}
with corresponding score contributions
$c^{(f)}_{\text{norm}} = \Delta\mathbf{Q}^{(f)}_{\text{norm}}
 \cdot \mathbf{K}^{(f)}_{\text{avg}}$ and
$c^{(f)}_{\text{dir}} = \Delta\mathbf{Q}^{(f)}_{\text{dir}}
 \cdot \mathbf{K}^{(f)}_{\text{avg}}$,
which are exact and additive by construction
($c^{(f)}_{\text{norm}} + c^{(f)}_{\text{dir}} = \Delta S_f$).

Because this decomposition depends on the choice of reference direction,
we computed it twice: once using the internal-position query $\mathbf{Q}^{(f)}_{\text{int}}$ as reference (decomposition A) and once
using the position-0 query $\mathbf{Q}^{(f)}_{\text{pos0}}$ as reference
(decomposition B), and averaged the two attributions. This symmetric average is analogous to the Shapley symmetry axiom: it removes the directional asymmetry inherent in any single-reference decomposition and ensures neither endpoint is privileged. The residual
$|\hat{c}^{(f)}_{\text{norm}} + \hat{c}^{(f)}_{\text{dir}}
  - \Delta S_f|$ was $< 10^{-6}$ across all bands and sequences,
confirming exact additivity of the averaged decomposition.

\paragraph{Per-band fractions and global aggregation.}
For each band $f$, the fractional norm attribution is defined as
$\phi^{(f)}_{\text{norm}} = \hat{c}^{(f)}_{\text{norm}} / \Delta S_f$
(and analogously for direction) averaged across sequences. Bands with small or sign-inconsistent
$|\overline{\Delta S_f}|$ ($f_3$, $f_4$, $f_5$) were excluded from mechanistic interpretation because small denominators render the fractional decomposition numerically unreliable. A global attribution
was computed as an absolute-value-weighted average across the remaining
bands:
\begin{equation}
    \Phi_{\text{norm}}
    = \sum_f w_f \, \phi^{(f)}_{\text{norm}},
    \quad
    w_f = \frac{|\overline{\Delta S_f}|}{\sum_{f'} |\overline{\Delta S_{f'}}|},
    \label{eq:global_phi}
\end{equation}
where the sum runs over all 8 bands including the excluded ones (their
small $|\overline{\Delta S_f}|$ values contribute negligible weights).
Signed weighting was not used because bands with negative
$\overline{\Delta S_f}$ would otherwise distort the aggregate. The
robustness of the global attribution to reference-direction choice was
assessed by reporting $\Phi_{\text{norm}}$ separately for decompositions
A and B alongside the symmetric average
(Table~\ref{tab:ext-rope-decomp}).

All computations were performed across $n = 57$ M-predicting sequences
with the masked token at position~0 (clean condition) and position~10
(internal condition).Position~10 was chosen as a representative internal position well clear of $f_0$'s $\sim 6$-token rotation period, ensuring that the decomposition reflects the typical case rather than positions where $f_0$ alignment is partially preserved by RoPE geometry.

\paragraph{Computational procedure.}
The decomposition was computed across L6H8's $8$ RoPE frequency bands
(head dimension $= 16$, non-interleaved format), separately for all
$n = 57$ M-predicting sequences with the masked token at position~0
(clean condition) and position~10 (internal condition). For each
frequency pair, the score difference between position-$0$ and
internal-position runs was decomposed into norm and direction
contributions. To avoid attributing the change asymmetrically to either
endpoint, each attribution was computed twice (once with the
internal-position query as reference direction, once with the
position-$0$ query) and the results averaged, analogous to the Shapley
symmetry axiom. Because the BOS key representation showed a small
$2$--$7\%$ positional variation in norm, the mean key vector across
both conditions was used as a neutral reference rather than privileging
either condition. The decomposition is exact by construction
(residual $< 10^{-6}$).

% =============================================================================
% EXTENDED DATA FIGURES
% =============================================================================
\clearpage
\appendix

\section{Extended Data Figures}

% --- Extended Data: Full L6 attention patterns ---
\begin{figure}[htbp!]
    \centering
    \includegraphics[width=\textwidth]{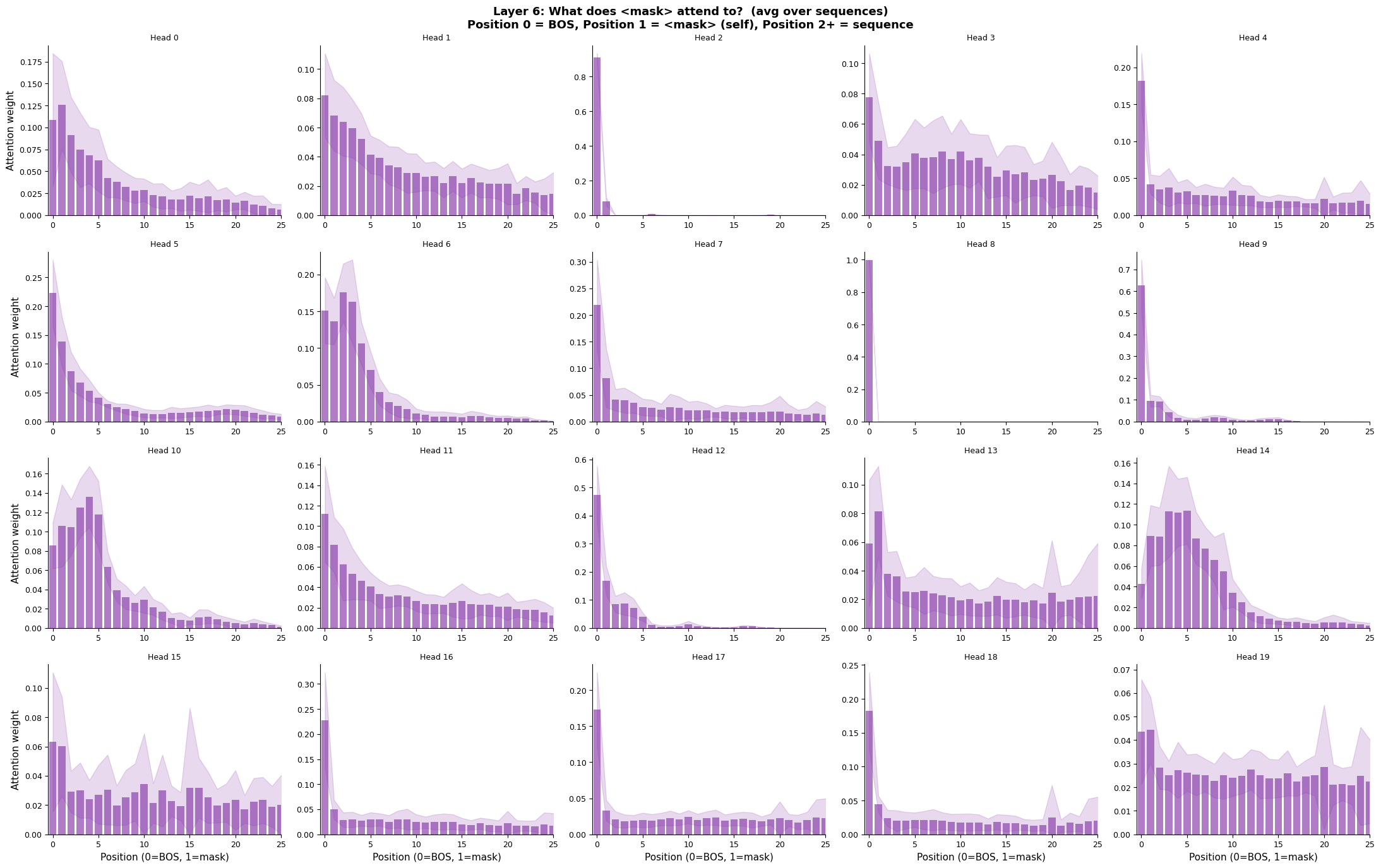}
    \caption{\textbf{Layer~6 attention patterns for all 20 heads.}
    Attention weights from the \texttt{<mask>} position (position~0 = BOS,
    position~1 = \texttt{<mask>}, position~2+ = sequence tokens), averaged
    over sequences. Each subplot shows one head's attention distribution.
    Only Head~8 shows near-exclusive BOS focus; other heads display diffuse
    attention or attend primarily to the MASK token itself or nearby
    sequence positions.}
    \label{fig:ext-l6-attention}
\end{figure}

% --- Extended Data: Full L6 attention patterns ---
\begin{figure}[htbp!]
    \centering
    \includegraphics[width=\textwidth]{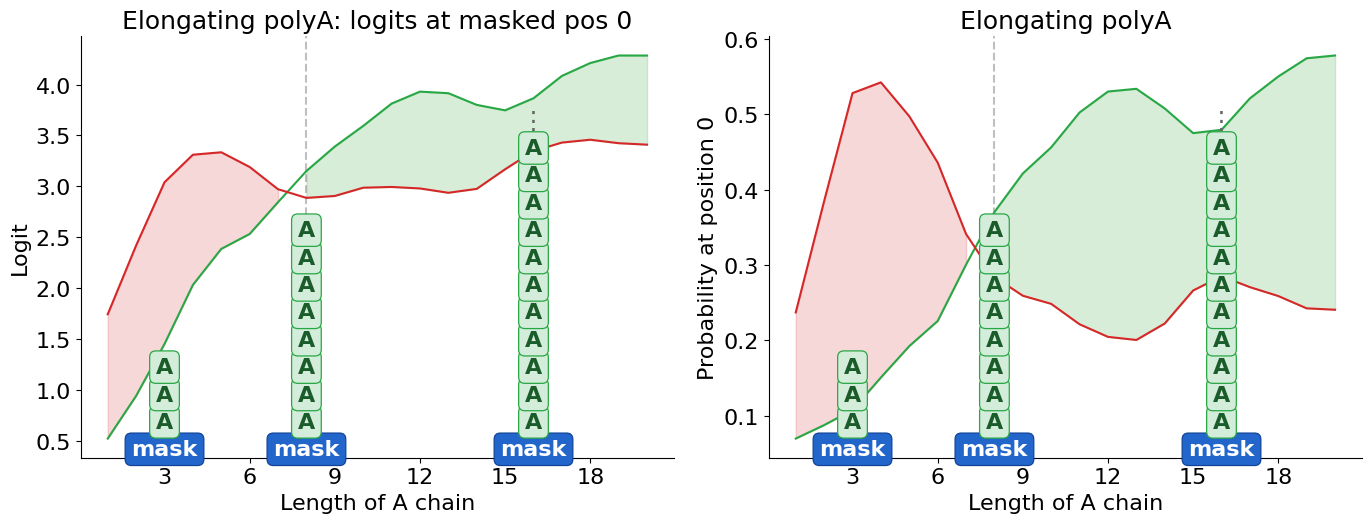}
    \includegraphics[width=\textwidth]{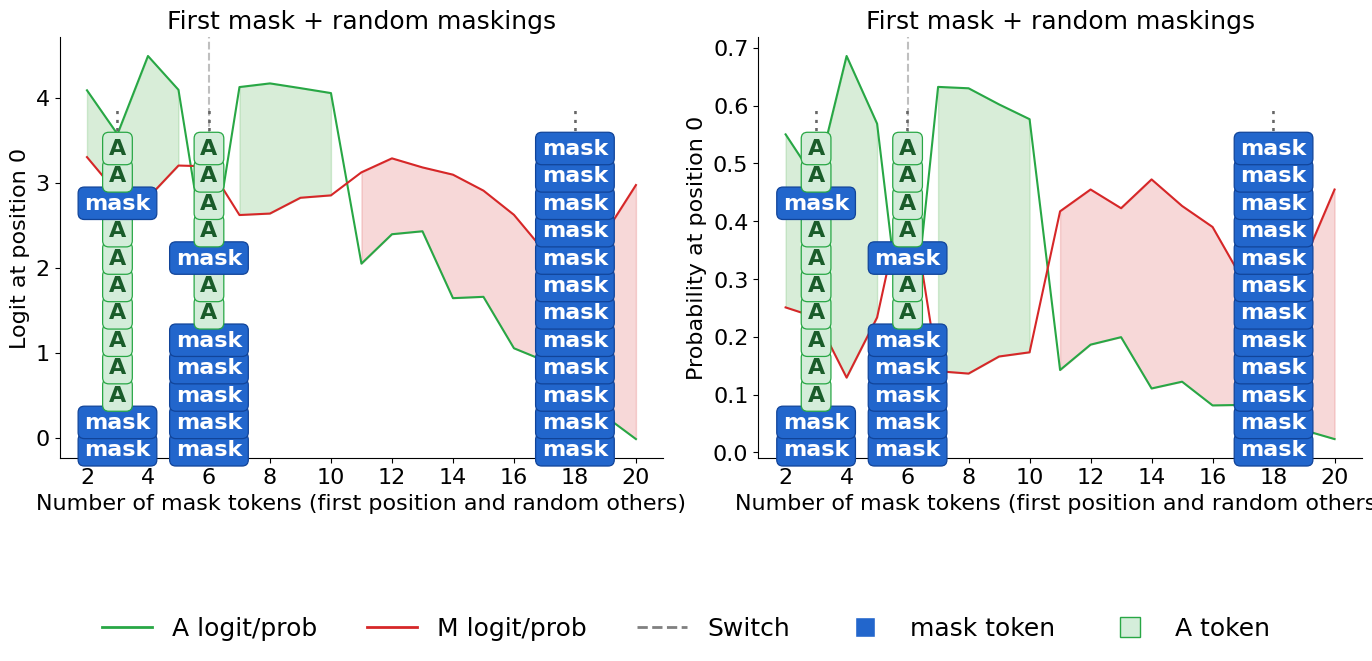}
    \caption{\textbf{Methionine prediction emerges from competition
    between positional and contextual circuits.}
    \textbf{(A,~B)}~Context elongation. A MASK at position~0 is followed
    by Poly-A chains of increasing length (1--20 residues). (A)~Logits
    for A and M at the masked position; the A logit rises while M remains
    stable, with a crossover (dashed ``Switch'' line) at $\sim$9--10
    residues. (B)Corresponding softmax probabilities.
    \textbf{(C,~D)}~Context destruction. Starting from a 20-residue Poly-A
    sequence with MASK at position~0, additional masks are introduced at
    random internal positions. (C)~Logits: the A logit decreases as
    context is destroyed while M remains constant; M re-emerges at
    $\sim$8--10 total masks. (D)~Corresponding probabilities.}
    \label{fig:ext-polyA-experiments}
\end{figure}

% --- Extended Data: BOS _vs_all_ablation ---
\begin{figure}[htbp!]
    \centering
    \includegraphics[width=\textwidth]{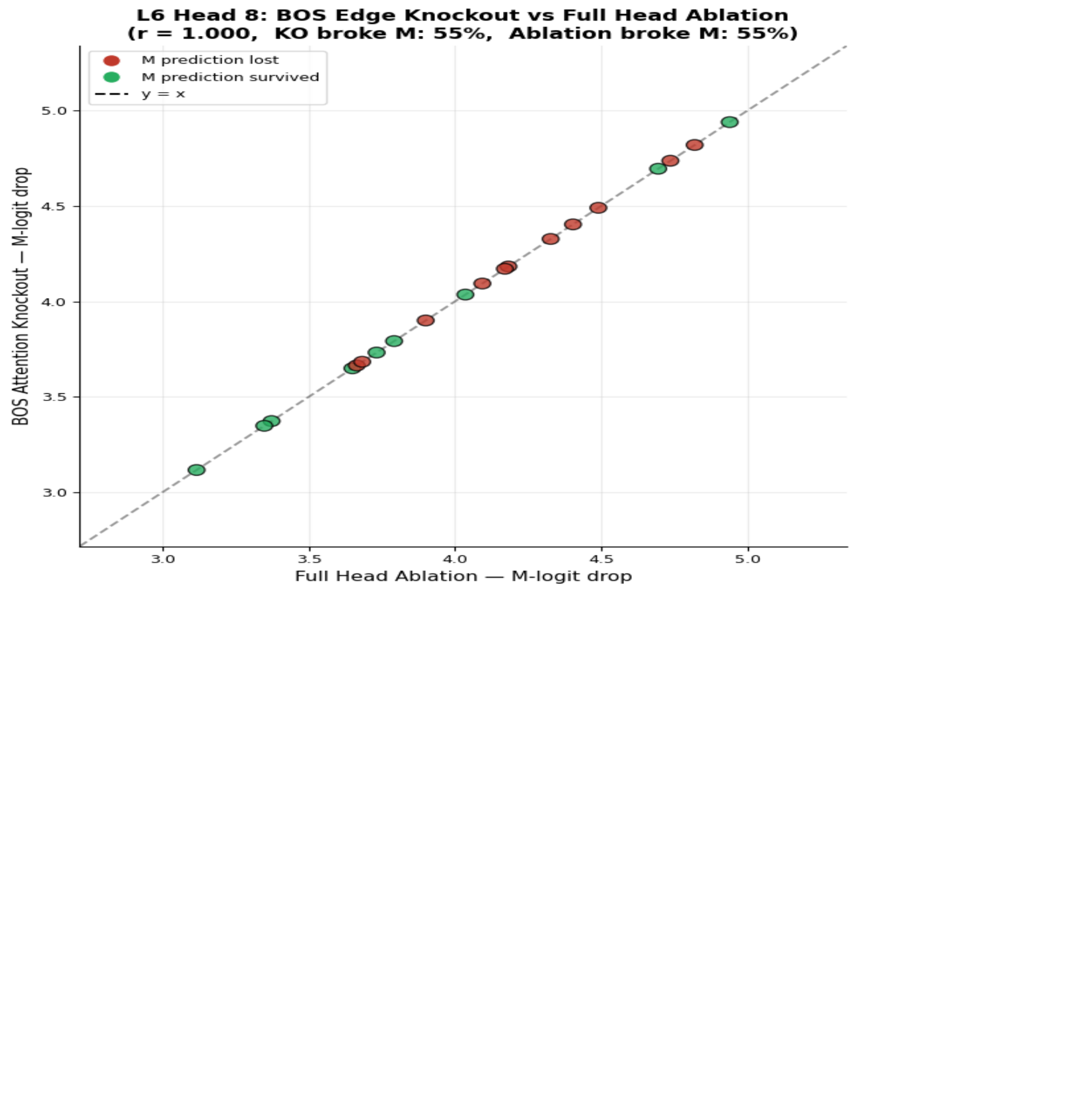}
    \caption{\textbf{BOS-edge knockout ($y$-axis) versus full
    head ablation ($x$-axis)}
    ~M-logit drop from BOS-edge knockout ($y$-axis) versus full
    head ablation ($x$-axis). Points lie on the identity line ($r = 1.0$),
    confirming that L6H8 operates via the BOS edge. Green = M prediction
    survived; red = M prediction lost.}
    \label{fig:ext-l6-bos}
\end{figure}
% --- Extended Data: rope basic abaltion ---
\begin{figure}[htbp!]
    \centering
    \includegraphics[width=\textwidth]{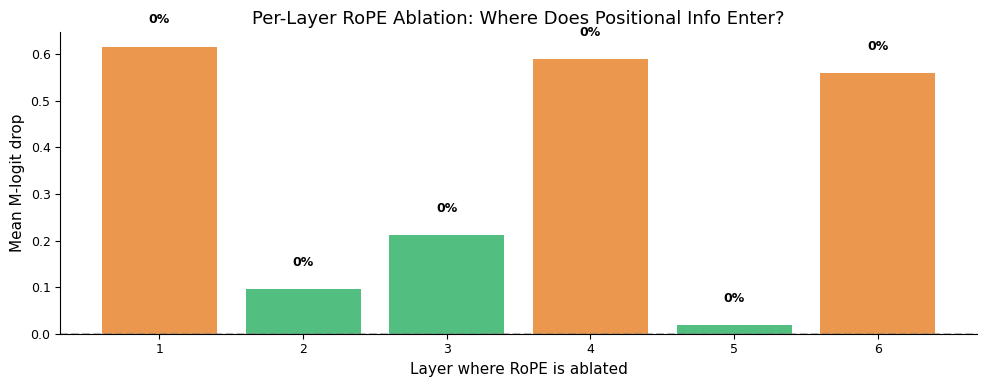}
    \caption{\textbf{RoPE ablations in each attention layer}
    the results show that RoPE signal is redundant between the different layer.}
    \label{fig:ext-basic-rope-ablation}
\end{figure}
% --- Extended Data: rope_cosines_and_norms ---
\begin{figure}[htbp!]
    \centering
    \includegraphics[width=0.8\textwidth]{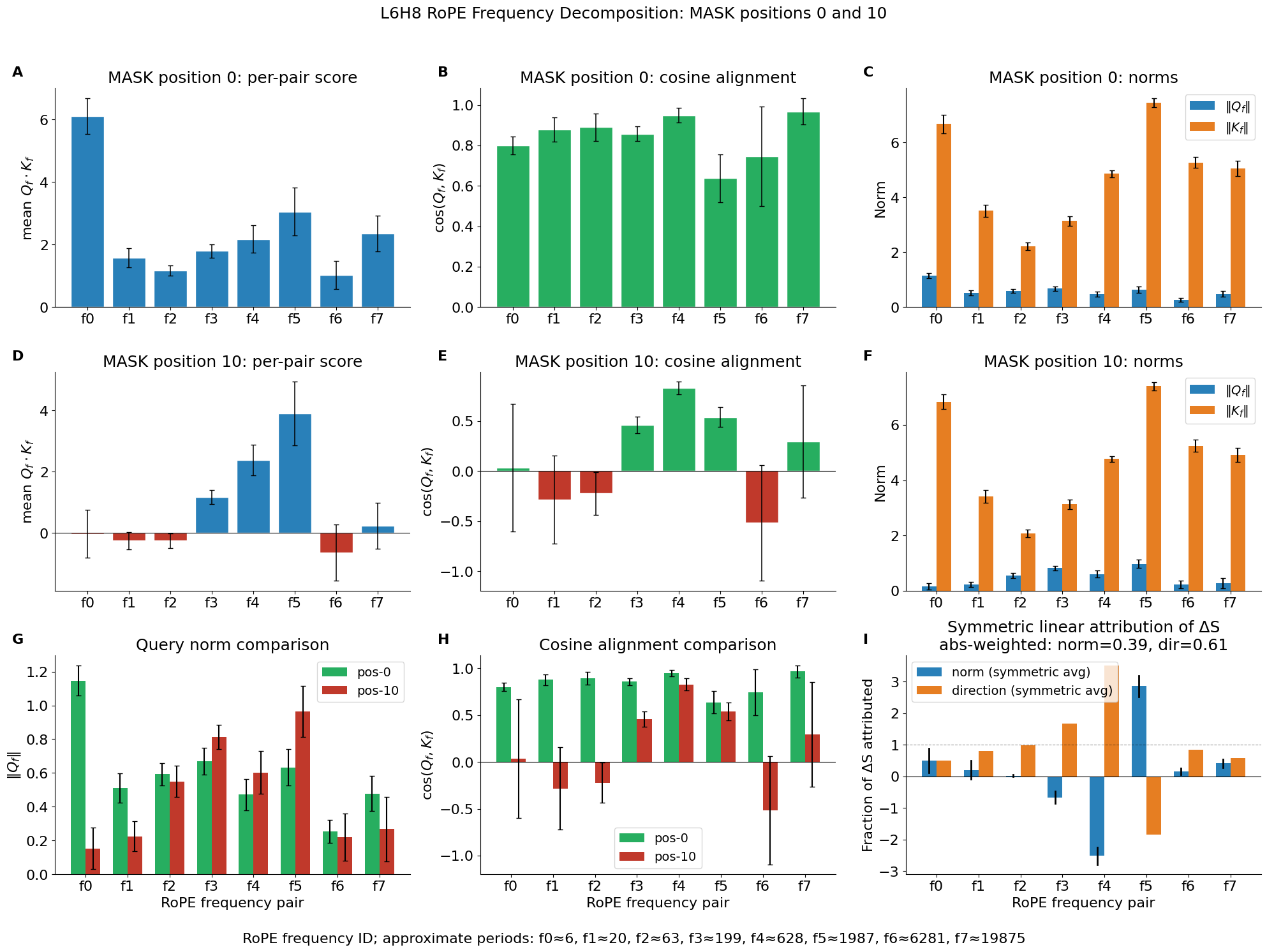}
    \caption{\textbf{RoPE frequency decomposition of L6H8's BOS attention and symmetric norm--direction attribution of the positional score difference.} 
    \label{fig:rope-freq-dec}
All panels show results across $n=57$ methionine-predicting sequences. 
\textbf{(A--C)}~Decomposition with \texttt{<MASK>} at position~0. 
\textbf{(A)}~Mean per-band dot product $\mathbf{Q}^{(f)}_{\text{MASK}} \cdot \mathbf{K}^{(f)}_{\text{BOS}}$ across 8 RoPE frequency pairs; $f_0$ is the dominant contributor ($\overline{S_{f_0}} = 6.08$). 
\textbf{(B)}~Cosine similarity $\cos(\mathbf{Q}^{(f)}, \mathbf{K}^{(f)})$ per frequency pair; values are uniformly high at position~0 ($f_0$: $\sim 0.80$), confirming strong angular alignment across all bands. 
\textbf{(C)}~Mean norms of $\mathbf{Q}^{(f)}$ (blue) and $\mathbf{K}^{(f)}$ (orange); BOS key norms are substantially larger than query norms at position~0. 
\textbf{(D--F)}~Same decomposition with \texttt{<MASK>} at internal position~10, chosen as a representative internal position well clear of $f_0$'s $\sim 6$-token rotation period. 
\textbf{(D)}~Per-band dot products are substantially reduced relative to position~0; $f_0$ collapses to near zero and $f_1, f_2, f_6$ become negative. 
\textbf{(E)}~Cosine similarities collapse across multiple bands: $f_0$ falls from $\sim 0.8$ to near zero, and $f_1, f_2, f_6$ flip sign. 
\textbf{(F)}~BOS key norms remain effectively position-invariant (compare with C); query norms show modest variation in most bands but a sharp drop in $f_0$. 
\textbf{(G)}~Direct comparison of query norms $\|\mathbf{Q}^{(f)}\|$ at position~0 (green) versus internal position~10 (red). The $f_0$ band shows a stark seven-fold difference (mean 1.14 vs.\ 0.15), while other bands differ only modestly, establishing that the dominant positional norm effect is concentrated in $f_0$ alongside a cosine collapse in the same band. 
\textbf{(H)}~Direct comparison of cosine similarities at position~0 versus internal position~10. The $f_0, f_1, f_2,$ and $f_6$ bands all show large directional drops, with $f_1, f_2, f_6$ flipping sign and $f_0$ collapsing to near zero --- consistent with their role as alignment selectors in the attribution analysis. 
\textbf{(I)}~Symmetric linear attribution of $\Delta S_f$ into norm (blue) and direction (orange) components per frequency band (see Methods). Black bars indicate the range across the two reference choices (decompositions A and B). The $f_0$ band engages both mechanisms approximately equally (50\% norm, 50\% direction), with the largest reference-choice spread of any band because both factors change substantially. Bands $f_1, f_2,$ and $f_6$ are predominantly direction-driven (80--98\%); $f_7$ is mixed ($\approx 60\%$ direction). Bands $f_3, f_4,$ and $f_5$ have small or negative $\overline{\Delta S_f}$ and show large, unreliable attribution fractions (italicised in Table~\ref{tab:ext-rope-decomp}). Global abs-weighted attribution: norm $= 0.39$, direction $= 0.61$ (Table~\ref{tab:ext-rope-decomp}).}
\end{figure}

% --- Extended Data: L6H8 per-sequence prediction ---
\begin{figure}[htbp!]
    \centering
    \includegraphics[width=\textwidth]{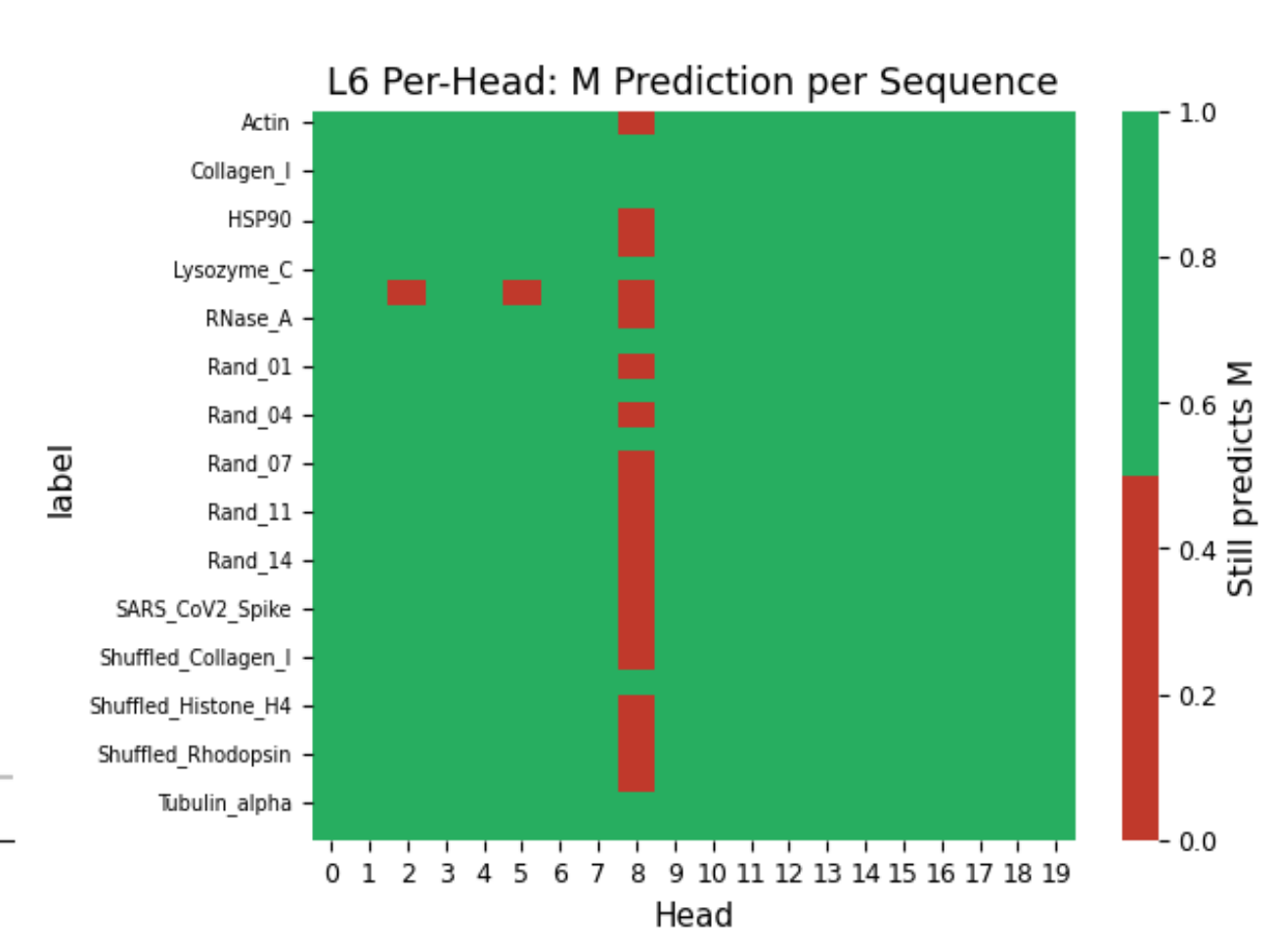}
    \caption{\textbf{Per-sequence M prediction upon Layer~6 head ablation.}
    Heatmap showing whether each sequence (rows) still predicts M as the
    top-1 token after ablating each layer ~6 head (columns 0--19). Green indicates methionine prediction survives; red indicates failure. L6H8 (Head~8) ablation causes near-universal failure across all sequence types, whereas ablation of other heads has minimal effect. A small number of sequences
    (e.g., RNase\_A) also lose M prediction upon ablation of Heads~6--7,
    suggesting minor secondary contributions.}
    \label{fig:ext-l6-perseq}
\end{figure}

% --- Extended Data: MLP positional role ---
\begin{figure}[htbp!]
    \centering
    \includegraphics[width=\textwidth]{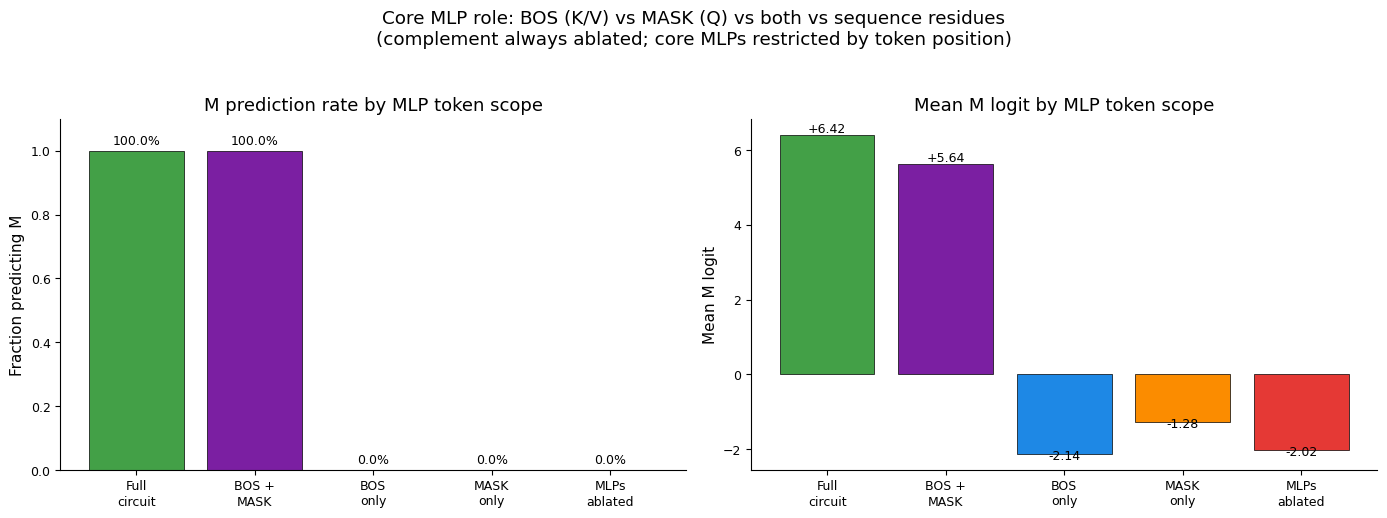}
    \caption{\textbf{MLPs must process both BOS and MASK tokens.}
    M-prediction rate (left) and mean M logit (right) under five MLP ablation scopes within the minimal circuit. ``Full circuit'': all
    core MLPs active (100\%, logit +6.42). ``BOS + MASK'': core MLPs
    active only at the BOS and MASK positions (100\%, +5.64). ``BOS only'':
    MLPs active only at BOS (0\%, $-$2.14). ``MASK only'': MLPs active
    only at MASK (0\%, $-$1.28). ``MLPs ablated'': all core MLPs removed
    (0\%, $-$2.02). Preserving MLPs at both positions is necessary and
    sufficient; either alone fails.}
    \label{fig:ext-mlp-scope}
\end{figure}

% --- Extended Data: MLP and Attention positional role ---
\begin{figure}[htbp!]
    \centering
    \includegraphics[width=\textwidth]{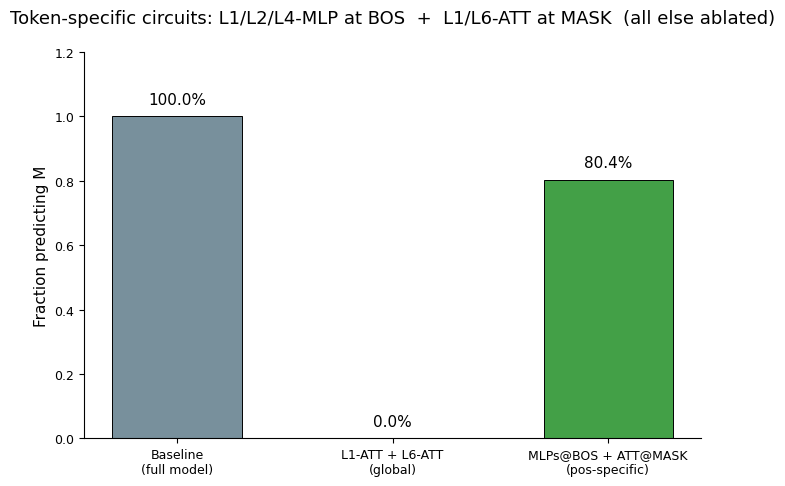}
    \caption{\textbf{Causal sufficiency test. }
    The full model predicts M at 100\%;
    L1+L6 attention alone (global ablation) drops to 0\%; restoring MLPs
    at BOS (L1/L2/L4) plus attention at the masked token (position-specific)
    rescues to 80.4\%.}
    \label{fig:ext-mlp-bos-attn-mask}
\end{figure}

% --- Extended Data: RoPE internal-position shift ---
\begin{figure}[htbp!]
    \centering
    \includegraphics[width=\textwidth]{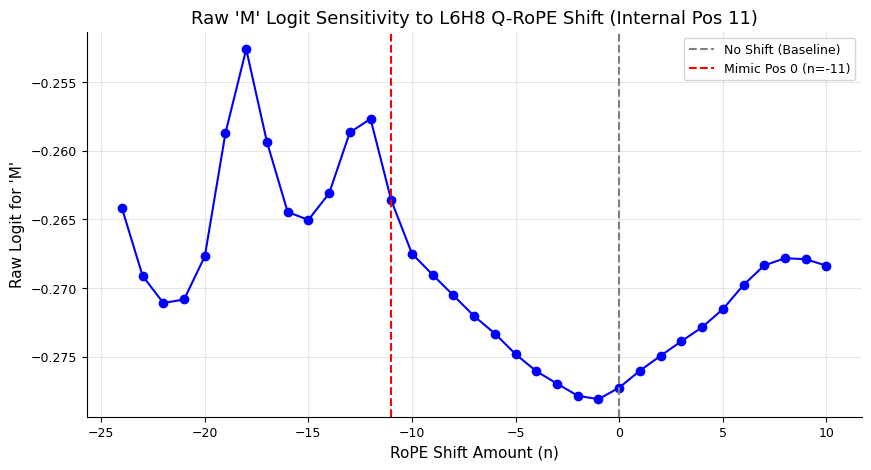}
    \caption{\textbf{Shifting L6H8's Q-RoPE to mimic position~0 at an
    internal position.} Raw M logit as a function of RoPE shift $n$ at
    internal position~11. A clear spike at $n = -11$ (red dashed line,
    ``Mimic Pos~0'') confirms that making L6H8 ``see'' position~0 boosts
    M prediction. The baseline (no shift, grey dashed) is shown for
    reference. The effect is modest in absolute terms because the rest
    of the model's residual stream still encodes the internal position
    identity.}
    \label{fig:ext-rope-shift}
\end{figure}

% --- Extended Data: Pre vs Post RoPE Q patching ---
\begin{figure}[htbp!]
    \centering
    \includegraphics[width=\textwidth]{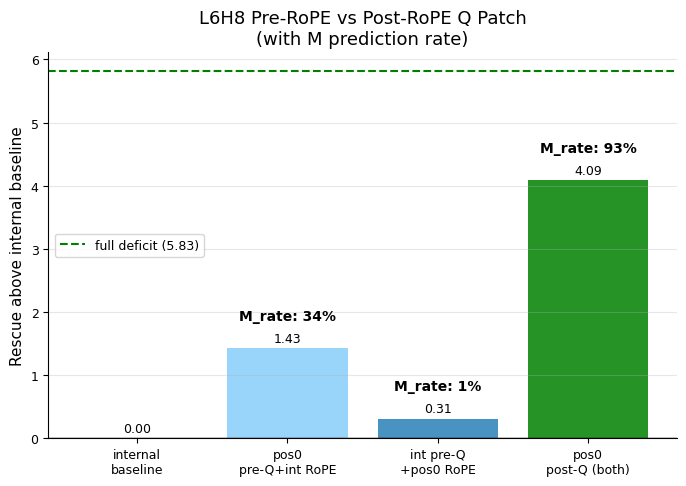}
    \caption{\textbf{Pre-RoPE versus post-RoPE query patching at L6H8.}
    Rescue above internal baseline (bar height) and M-prediction rate
    (labels) for four conditions. ``Internal baseline'': no patching
    (M-rate 0\%, rescue 0.00). ``pos0 pre-Q + int RoPE'': position-0
    pre-RoPE query content combined with internal-position RoPE rotation
    (M-rate 34\%, rescue 1.43). ``int pre-Q + pos0 RoPE'': internal
    pre-RoPE content with position-0 RoPE (M-rate 1\%, rescue 0.31).
    ``pos0 post-Q (both)'': full position-0 post-RoPE query (M-rate 93\%,
    rescue 4.09). Green dashed line shows the full deficit (5.83).
    Neither pre-RoPE content nor RoPE rotation alone is sufficient;
    together they provide near-complete rescue.}
    \label{fig:ext-qk-permutation}
\end{figure}

% --- Extended Data: Cross-sequence activation injection ---
\begin{figure}[htbp!]
    \centering
    \includegraphics[width=0.7\textwidth]{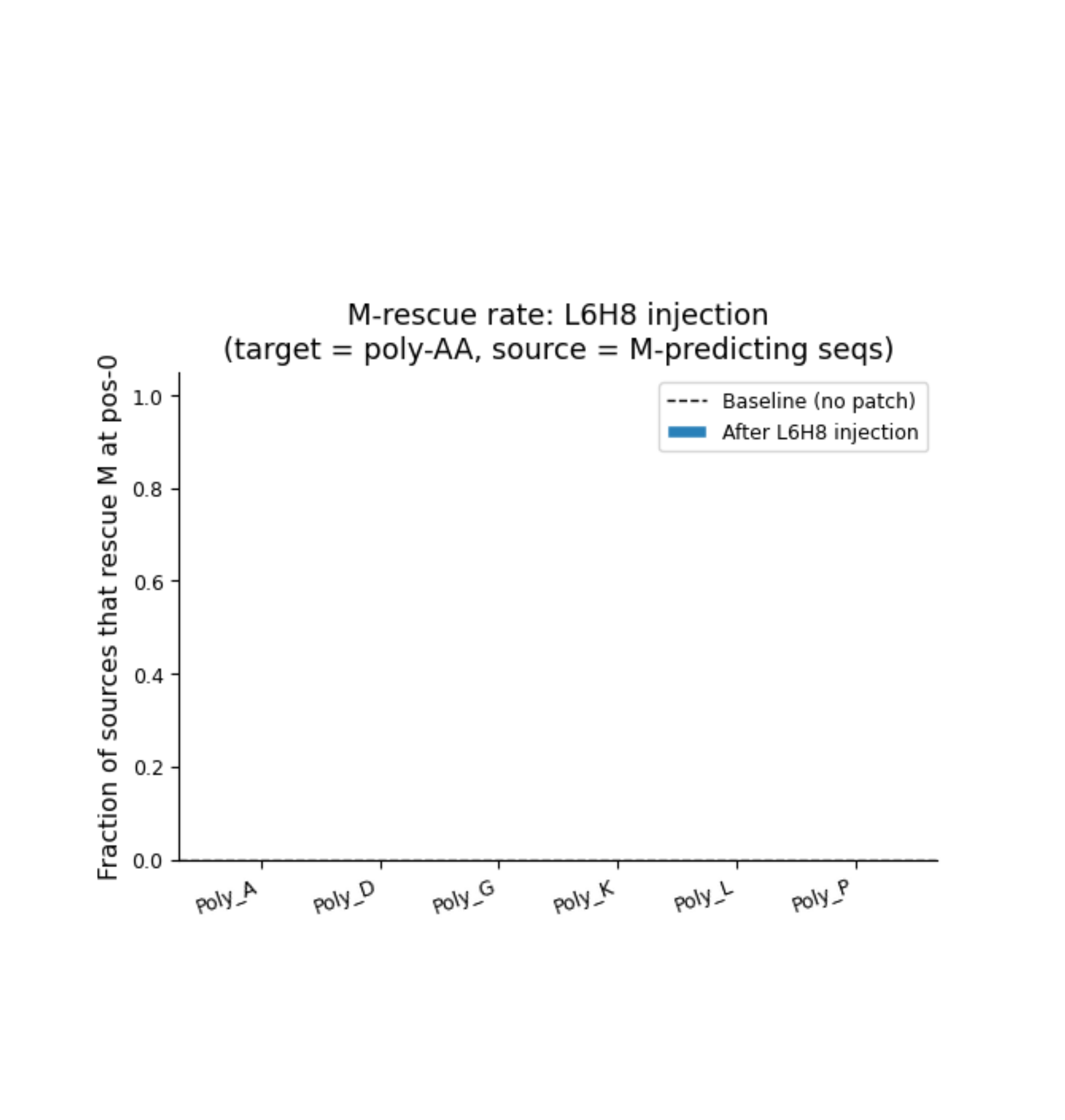}
    \caption{\textbf{Cross-sequence activation injection does not rescue M
    prediction in poly-amino-acid targets.} L6H8 pre-projection activations
    from M-predicting source sequences are transplanted into poly-amino-acid
    targets (Poly-A, Poly-D, Poly-G, Poly-K, Poly-L, Poly-P). The injection
    does not rescue M prediction in any target, demonstrating that competing
    identity circuits overwhelm the M signal even when the methionine
    circuit's output is artificially strengthened.}
    \label{fig:ext-injection}
\end{figure}

\FloatBarrier
\clearpage

% =============================================================================
% EXTENDED DATA TABLES
% =============================================================================

\section{Extended Data Tables}

% --- Extended Table: Greedy head search ---
\begin{table}[htbp!]
    \centering
    \small
    \caption{Greedy RoPE-swap search identifies 12 key upstream
    heads. Heads are ordered by cumulative rescue. Swapping position-0
    RoPE into these heads during internal-MASK runs progressively
    recovers M prediction.}
    \label{tab:ext-greedy-heads}
    \begin{tabular}{clcc}
        \toprule
        Step & Head & M-rate rescue (\%) & Logit Rescue (\%) \\
        \midrule
        1  & L5H9  & 6.3  & 8.4  \\
        2  & L1H8  & 20.7 & 19.6 \\
        3  & L4H3  & 40.7 & 32.8 \\
        4  & L5H17 & 58.6 & 41.4 \\
        5  & L5H2  & 71.6 & 47.8 \\
        6  & L5H7  & 74.4 & 49.5 \\
        7  & L1H9  & 75.8 & 51.3 \\
        8  & L1H1  & 77.5 & 52.2 \\
        9  & L4H19 & 78.2 & 52.8 \\
        10 & L1H18 & 78.6 & 53.7 \\
        11 & L4H7  & 78.9 & 54.1 \\
        12 & L1H5  & 79.3 & 55.6 \\
        \bottomrule
    \end{tabular}
\end{table}

% --- Extended Data Table: RoPE expansion baselines ---
\begin{table}[htbp!]
    \centering
    \small
    \caption{Baselines and all-head expansion for positional rescue. Values
    represent M-prediction rescue when swapping position-0 RoPE into
    progressively more heads.}
    \label{tab:ext-rope-expansion}
    \begin{tabular}{lcccl}
        \toprule
        Condition & M-gap & M-rate & Rescue (\%) & Description \\
        \midrule
        pos-0 clean & 3.73 & 1.000 & 100.0 & Upper reference \\
        Internal baseline & $-$2.10 & 0.000 & 0.0 & Lower reference \\
        L6H8 full-Q patch & 1.99 & 0.933 & 70.2 & Theoretical max \\
        L6H8 RoPE only & $-$2.01 & 0.007 & 1.5 & L6H8 alone \\
        3 key heads + L6 & $-$0.19 & 0.407 & 32.8 & L1H8, L4H3, L5H9 \\
        L1+L4+L5 all + L6 & 1.99 & 0.888 & 70.1 & 60 heads \\
        \bottomrule
    \end{tabular}
\end{table}

% --- Extended Data Table: RoPE frequency contributions ---
\begin{table}[htbp!]
    \centering
    \small
    \caption{RoPE frequency contributions to L6H8's BOS attention.
    Per-frequency-pair contribution to the
    $\mathbf{Q}_\text{MASK} \cdot \mathbf{K}_\text{BOS}$ dot product,
    averaged over M-predicting sequences.}
    \label{tab:ext-rope-freq}
    \begin{tabular}{cccc}
        \toprule
        Frequency ($f$) & Angular frequency ($\omega_f$) &
        Period (tokens) & Contribution (\%) \\
        \midrule
        0 & 1.000 & 6.28 & 31.7 \\
        5 & 0.003 & 1{,}988 & 15.9 \\
        7 & 0.0003 & 19{,}635 & 12.2 \\
        4 & 0.010 & 628 & 11.3 \\
        3 & 0.032 & 199 & 9.3 \\
        1 & 0.316 & 20 & 8.2 \\
        2 & 0.100 & 63 & 6.1 \\
        6 & 0.001 & 6{,}283 & 5.3 \\
        \bottomrule
    \end{tabular}
\end{table}

% --- Extended Data Table: RoPE decomposition ---

\begin{table}[htbp!]
    \centering
    \small
    \caption{Symmetric norm--direction decomposition of L6H8's positional
    score difference $\Delta S_f = \Delta\mathbf{Q}^{(f)} \cdot
    \mathbf{K}^{(f)}_{\text{avg}}$ across RoPE frequency bands.
    $\Delta S_f$ is the mean per-band score difference (position-0 minus
    internal, $n = 57$ M-predicting sequences). The abs-weight column gives
    fractional contribution of each band to the total absolute $\Delta S$.
    Norm and direction fractions report the symmetric linear attribution
    (average of attributions computed with internal-position and
    position-0 queries as reference direction; see Methods). Bands with small or sign-inconsistent
    $\Delta S_f$ ($f_3$, $f_4$, $f_5$) have unreliable fractional
    decompositions and are italicised. Global (absolute-weighted) attribution:
    norm $= 0.385$ (range 0.153--0.618 across reference choices),
    direction $= 0.615$.}
    \label{tab:ext-rope-decomp}
    \begin{tabular}{cccccc}
        \toprule
        Band & $\omega_f$ & $\overline{\Delta S_f}$ & Abs-weight & 
        Frac.\ norm & Frac.\ direction \\
        \midrule
        $f_0$ & 1.000  &  6.219 & 0.419 & 0.501 & 0.499 \\
        $f_1$ & 0.316  &  1.816 & 0.122 & 0.203 & 0.797 \\
        $f_2$ & 0.100  &  1.390 & 0.094 & 0.019 & 0.981 \\
        $f_6$ & 0.001  &  1.676 & 0.113 & 0.159 & 0.841 \\
        $f_7$ & 0.0003 &  2.070 & 0.139 & 0.409 & 0.591 \\
        \midrule
        \emph{$f_3$} & \emph{0.032} & \emph{0.661}  & \emph{0.044} & 
            \emph{$-0.668$} & \emph{1.668} \\
        \emph{$f_4$} & \emph{0.010} & \emph{$-0.257$} & \emph{0.017} & 
            \emph{$-2.520$} & \emph{3.520} \\
        \emph{$f_5$} & \emph{0.003} & \emph{$-0.766$} & \emph{0.052} & 
            \emph{2.856}  & \emph{$-1.856$} \\
        \bottomrule
    \end{tabular}
\end{table}
% =============================================================================
% REFERENCES
% =============================================================================
\bibliography{references}
\bibliographystyle{unsrtnat}

\section{acknowledgments}
PJ is supported by funding from the Biotechnology and Biological Sciences Research Council UKRI-BBSRC grant (BB/T008784/1). PJ cofounded Evolvere Biosciences, but the company had no role in this study. OMC acknowledges funding from a New College Todd-Bird Junior Research Fellowship and MRC Fellowship MR/Y010078/1 . He acknowledges consulting fees to Pelago Biosciences, Faculty.ai, MarketCast and is on the scientific advisory board of Evolvere Biosciences. No funder had a role in the research or decision to publish

\end{document}